\PassOptionsToPackage{table,xcdraw}{xcolor}
\documentclass[conference,compsoc]{IEEEtran}
\usepackage{cite}
\usepackage{amsmath,amssymb,amsfonts}
\usepackage{graphicx}
\usepackage{textcomp}
\def\BibTeX{{\rm B\kern-.05em{\sc i\kern-.025em b}\kern-.08em
    T\kern-.1667em\lower.7ex\hbox{E}\kern-.125emX}}
\PassOptionsToPackage{table,xcdraw}{xcolor}

\usepackage{microtype}
\microtypecontext{spacing=nonfrench}
\usepackage{tikz}
\usepackage{amsmath}
\usepackage{comment}
\usepackage{mathtools}
\usepackage{booktabs}
\usepackage{lipsum}
\usepackage{listings}
\usepackage{xspace}
\usepackage{listings}
\usepackage{bbding}
\usepackage{pifont}
\usepackage{wasysym}
\usepackage{amssymb}
\usepackage{amsmath}
\usepackage{amsfonts}
\usepackage{amssymb}
\usepackage{caption}
\usepackage{setspace}
\usepackage{tcolorbox}
\usepackage{float}
\usepackage[table]{xcolor}
\usepackage{color}
\definecolor{Gray}{gray}{0.9}
\usepackage{algorithm}
\usepackage{algpseudocode}
\usepackage{multirow}
\usepackage{hyperref}
\usepackage{cleveref}
\usepackage{fancyhdr}
\usepackage[utf8]{inputenc}
\usepackage{amsfonts}
\usepackage{amssymb}
\usepackage{graphicx}
\usepackage[symbol]{footmisc}
\usepackage{enumitem}
\usepackage{tikz}

\ifCLASSOPTIONcompsoc
 \usepackage[caption=false,font=footnotesize,labelfont=sf,textfont=sf]{subfig}
\else
 \usepackage[caption=false,font=footnotesize]{subfig}
\fi

\makeatletter
\def\RemoveSpaces#1{\zap@space#1 \@empty}
\makeatother

\newcommand*\halfcirc[1][1ex]{%
  \begin{tikzpicture}
  \draw[fill=green] (0,0)-- (90:#1) arc (90:270:#1) -- cycle ;
  \draw (0,0) circle (#1);
  \end{tikzpicture}}
\newcommand{\xmark}{\ding{55}}%
\definecolor{verde}{rgb}{0.25,0.5,0.35}
\definecolor{myyellow}{RGB}{229, 148, 12}
\definecolor{jpurple}{rgb}{0.5,0,0.35}
\definecolor{darkgreen}{rgb}{0.0, 0.2, 0.13}
\definecolor{oldmauve}{rgb}{0.4, 0.19, 0.28}
\definecolor{blueannoback}{RGB}{234,242,250}
\newcommand{\mycode}{
\lstset{
    language=Java,
    basicstyle=\ttfamily\small,
    keywordstyle=\color{blue}\bfseries,
    keywordstyle=[2]\color{blue}\bfseries,
    keywordstyle=[3]\color{jpurple}\bfseries,
    deletekeywords = {return, if, public, private},
    morekeywords=[2]{function, public, private, internal, external, balanceOf},
    morekeywords=[3]{address, IERC20, uint, msg.sender, to, from},
    escapeinside={\%*}{*)},  
    stringstyle=\color{oldmauve},
    commentstyle=\color{red},
    morecomment=[s][\color{darkgreen}]{/*}{*/},
    morecomment=[s][\color{red}]{@}{;},
    extendedchars=true,
    showspaces=false,
    showstringspaces=false,
    numbers=left,
    numberstyle=\tiny,
    breaklines=true,
    breakautoindent=true,
    captionpos=b,
    xleftmargin=0pt,
    tabsize=2,
    float,
    floatplacement=!t,
    numbersep=5pt
}}

\newcommand{\ie}{i.e., }
\newcommand{\eg}{e.g., }
\newcommand{\corral}{\textsc{Corral} \cite{Corral}}

\newcommand{\smartest}{\textsc{SmarTest} \cite{smartest}}
\newcommand{\smartestnq}{\textsc{SmarTest}}
\newcommand{\slither}{\textsc{Slither} \cite{slither}}
\newcommand{\slithernq}{\textsc{Slither}}

\newcommand{\manticore}{\textsc{Manticore} \cite{manticore}}
\newcommand{\manticorenq}{\textsc{Manticore}}

\newcommand{\mythril}{\textsc{Mythril} \cite{mythril}}
\newcommand{\mythrilnq}{\textsc{Mythril}}

\newcommand{\verisol}{\textsc{VeriSol} \cite{verisol}}
\newcommand{\verisolnq}{\textsc{VeriSol}}

\newcommand{\verisolsnq}{\textsc{VeriSol}'s}
\newcommand{\verismart}{\textsc{VeriSmart} \cite{verismart}}
\newcommand{\verismartnq}{\textsc{VeriSmart}}
\newcommand{\solc}{\textsc{Solc-Verify} \cite{solc}}

\newcommand{\zeus}{\textsc{Zeus} \cite{zeus}}

\newcommand{\invcon}{\textsc{InvCon} \cite{invcon}}
\newcommand{\invconnq}{\textsc{InvCon}}

\newcommand{\daikon}{\textsc{Daikon} \cite{daikon}}
\newcommand{\daikonnq}{\textsc{Daikon}}

\newcommand{\maian}{\textsc{Maian} \cite{suicidalcon}}
\newcommand{\maiannq}{\textsc{Maian}}
\newcommand{\teether}{\textsc{TeEther} \cite{teether}}
\newcommand{\teethernq}{\textsc{TeEther}}
\newcommand{\ethbmc}{\textsc{EthBMC} \cite{ethbmc}}
\newcommand{\ethbmcnq}{\textsc{EthBMC}}
\newcommand{\oyente}{\textsc{Oyente} \cite{smart}}
\newcommand{\osiris}{\textsc{Osiris} \cite{osiris}}
\newcommand{\honeybadger}{\textsc{HoneyBadger} \cite{honeypot}}
\newcommand{\confuzzius}{\textsc{ConFuzzius} \cite{confuzzius}}
\newcommand{\echidna}{\textsc{Echidna} \cite{echidna}}
\newcommand{\fluffy}{\textsc{Fluffy} \cite{diffuzz}}
\newcommand{\sfuzz}{\textsc{sFuzz} \cite{sfuzz}}
\newcommand{\svchecker}{\textsc{SVChecker} \cite{svchecker}}
\newcommand{\contractgraph}{\textsc{Neural Contract Graph} \cite{graphneural}}
\newcommand{\escort}{\textsc{ESCORT} \cite{neucheck}}
\newcommand{\lastsys}{\textsc{Eth2Vec} \cite{eth2vec}}
\newcommand{\smtchecker}{\textsc{SMTChecker} \cite{smt}}
\newcommand{\cider}{\textsc{Cider} \cite{learninginvariants}}

\newcommand{\chatgpt}{\textsc{ChatGPT}}
\newcommand{\gptscan}{\textsc{GPTScan}}
\newcommand{\ma}{\textsc{Manual\_Audit?}}

\def\sys{\textsc{SmartInv}\xspace}
\begin{document}\sloppy

\date{}
\title{\Large \bf \sys: Multimodal Learning for Smart Contract Invariant Inference}

\author{Sally Junsong Wang$^*$, Kexin Pei$^*$$^{\dagger}$, Junfeng Yang$^*$
\\  $^*$Columbia University, NY \xspace
    $^{\dagger}$The University of Chicago, IL\\
{jw4074@columbia.edu,}
kpei@cs.uchicago.edu, junfeng@cs.columbia.edu}

\maketitle



\begin{abstract}

Smart contracts are software programs that enable diverse business activities on the blockchain. Recent research has identified new classes of "machine un-auditable" bugs that arise from both transactional contexts and source code. Existing detection methods require human understanding of underlying transaction logic and manual reasoning across different sources of context (\ie modalities), such as code, dynamic transaction executions, and natural language specifying the expected transaction behavior.

To automate the detection of ``machine un-auditable'' bugs, we present \sys, an accurate and fast smart contract invariant inference framework. Our key insight is that the expected behavior of smart contracts, as specified by invariants, relies on understanding and reasoning across \emph{multimodal} information, such as source code and natural language. We propose a new prompting strategy to foundation models, Tier of Thought (ToT), to reason across multiple modalities of smart contracts and ultimately to generate invariants. By checking the violation of these generated invariants, \sys can identify potential vulnerabilities.

We evaluate \sys on real-world contracts and re-discover bugs that resulted in multi-million dollar losses over the past 2.5 years (from January 1, 2021 to May 31, 2023). Our extensive evaluation shows that \sys generates (3.5$\times$) more bug-critical invariants and detects (4$\times$) more critical bugs compared to the state-of-the-art tools in significantly (150$\times$) less time. \sys uncovers 119 zero-day vulnerabilities from the 89,621 real-world contracts. Among them, five are critical zero-day bugs confirmed by developers as ``high severity.''

\end{abstract}

\section{Introduction}
\label{sec:intro}

Bugs in smart contracts are often serious vulnerabilities that lead to significant loss of funds. 
In 2022 alone, \$2 billion was lost due to smart contract bugs \cite{PolyNetwork, DemystifyingBugs}. What makes smart contract bugs particularly damaging is the fact that once a smart contract is deployed, it becomes immutable, making it difficult to fix any vulnerabilities in the code.

Recent research has identified a new category of bugs known as ``machine un-auditable'' \emph{functional} bugs. 
These bugs, as their name suggests, cannot be reliably detected using existing automated tools that rely on pre-defined bug patterns \cite{DemystifyingBugs}. 
Unlike implementation bugs (\eg integer overflows), which often exhibit universal patterns that can be easily checked, functional bugs arise from a failure to reason about extensive domain-specific properties, \eg a specific transaction context. 
Detecting functional bugs require nontrivial reasoning across multiple sources of information or \emph{modalities}, such as source code and domain-specific rules expressed in natural language documentation. 

Unfortunately, traditional smart contract analyses rely heavily on manually-defined specifications by human experts~\cite{smt, solc, verismart, svchecker} and lack the ability to reason across \emph{multiple modalities}~\cite{smartest, ethbmc, teether, sfuzz, echidna}.
As a result, existing approaches are often tailored for specific types of bugs and do not generalize to new types of bugs.
Moreover, manual analysis is expensive, and thus not scalable to a large number of programs. 
While some tools \cite{CB} can find patterns in curated transactions of a limited class of functional bugs, none reliably detects functional bugs from source code due to functional bugs' diverse behaviour. Yet source code level bug detection is critical for preventing financial losses before contract deployment.
Functional bugs are also prevalent, accounting for 81\% exploitable bugs to date according to our survey (Table~\ref{table1}). 

\begin{table}[!t]

\small
\setlength{\tabcolsep}{6pt}
\centering
\renewcommand{\arraystretch}{1.1}

\caption{Statistics on Bountied Vulnerabilities of Solidity-based Smart Contracts (from September, 2021 to May, 2023)}
\label{table1}

\begin{tabular}{llll}
\toprule
Implementation & Functional & Others & Total \\
\midrule
929 (17.52\%) & 4,305 (81.20\%) & 68 (1.28\%) & 5,302\\
\bottomrule
\end{tabular}
\end{table}

Listing~\ref{example:multi-modal} shows an example of a functional bug. 
The \texttt{getPrice()} function computes \texttt{price} by the ratio of \texttt{token0} and \texttt{token1} values in \texttt{address(this)}. 
Ideally, \texttt{price} is expected to remain stable within a range.
However, as \texttt{token0} and \texttt{token1} are state variables, \ie static variables, they can be easily manipulated by external parties.
Therefore, when \texttt{getPrice()} is invoked to return \texttt{price}, its return value can fluctuate significantly, leading to the potential exploit of unexpected arbitrage trade on top of the newly manipulated price difference.
The functional bug of \texttt{getPrice()} stems from not effectively meeting transactional requirements.
For example, an outsized increase of \texttt{token1}'s balance value via another malicious contract can throw \texttt{price} off originally intended range (see \S\ref{subsec:motivating_examples} for details). Such functional bugs cannot be detected by existing automated tools that target low-level bugs, because the code shown does not contain implementation bugs.

\mycode
\begin{lstlisting}[float,floatplacement=H,basicstyle=\fontsize{8}{9}\selectfont, caption={functional bug example. An attacker can pump up \texttt{price} by inflating \texttt{token1}'s balance value.}, label={example:multi-modal}, numbers=left, xleftmargin=2em]
uint price;
IERC20 token0, token1;

function getPrice() {
  //functional bug: price manipulation
  price = token1.balanceOf(address(this))/token0.balanceOf(address(this));
}
\end{lstlisting}

We present \sys, an automated, foundation model based framework to infer smart contract invariants and to detect bugs at scale. While there are machine learning approaches for generating invariants \cite{cln2inv, mlinvariants, discoveringmlinvariants, learninginvariants,cider}, they predate foundation models, and thus follow the typical paradigm in hand-engineering limited features that can be helpful for contract invariant inference. The unique feature of \sys, which differentiates from existing analyzers, is that \sys leverages foundation models to reason about multimodal inputs, such as source code and natural language (comments and documentations on domain-specific transactional contexts). Foundation models are particularly suited for analyzing multimodal information and domain-specific bug detection, because they are pretrained on natural language texts and code, and can be further finetuned for domain-specific knowledge.

To reason across multimodal information, we develop \emph{Tier of Thought (ToT)}, a new prompting strategy, that can be used to finetune and elicit explicit reasoning of foundation models on the program structures of smart contract. 
In contrast to other foundation-model-based approaches \cite{gptscan, chatgpthowfar, manualaudit}, ToT applies universally across contract types, eliminating the need for bug-specific reasoning heuristics.
ToT breaks down the process of invariants generation into intermediate abstract tiers, based on the typical reasoning steps that human analyzers would take, such as predicting critical program points to check invariants, generating invariants associated with the predicted program points, and ranking the invariants by predicting their likelihood of preventing bugs. 
Based on the ranked invariants, \sys can efficiently verify the invariants by prioritizing the invariants that are more likely bug-preventive using a bounded model checker, without exhaustively enumerating all of them. 

Building \sys was an engineering effort to unite the potential of finetuning foundation models with the soundness of formal verification. We do not claim that multimodal learning and prompting are superior to traditional static and dynamic analyses, or vice versa. The goal of this paper is to explain \sys's design and implementation regarding how it overcomes the challenges posed by functional bugs, annd hopefully to present one promising way forward. 

\vspace{.1cm}\noindent\textbf{Contributions.} We make the following contributions:

\begin{itemize}[leftmargin=*]
\item To our best knowledge, \sys proposes the first finetuning approach that can both infer invariants and detect bugs by reasoning across multiple smart contract modalities, critical to detecting functional bugs pre-deployment.

\item 
We design a new prompting approach to foundation models, Tier of Thought (ToT), to finetune and elicit their reasoning by following the thought process of human analyzers, significantly improving the accuracy of the generated invariants and detected bugs while reducing runtime overhead. 

\item 
We implement our approach in \sys and demonstrate that \sys outperforms prior tools in invariants generation and functional bug detection. Notably, \sys has found 119 zero-day bugs in the wild. Among them, five are confirmed by the developers as high severity.

\item 
We collect a large (2,173 samples) annotated smart contract invariant dataset for training and 89,621 real-world contracts for bug detection. We release the datasets and the tool for public use at https://github.com/columbia/SmartInv.
\end{itemize}

\begin{table}[!t]
\footnotesize
\setlength{\tabcolsep}{0.8pt}
\centering
\renewcommand{\arraystretch}{1.1}

\caption{\sys detected bug classification. Modalities Required: the minimum modalities required to detect a given bug class. SC: source code. NL: natural langauge. Pattern Detectable: a bug class can be detected by identifying general patterns. ToT Detectable: a bug class can be detected by Tier of Thought. 
\halfcirc: a modality is sometimes required.}
\label{bugtab}

\begin{tabular}{lcccccc}
\toprule
 & \multicolumn{2}{c}{\bf Modalities} & \multirow{2}{*}{\bf \begin{tabular}[c]{@{}c@{}}Pattern\\ Detectable\end{tabular}} & \multirow{2}{*}{\bf \begin{tabular}[c]{@{}c@{}}ToT\\ Detectable\end{tabular}} \\
 & SC & NL   &  &  \\  

\midrule
\multicolumn{4}{l}{\bf Implementation Bugs\vspace{.1cm}}  \\
Reentrancy (RE) \cite{RE} & \color{green}\CheckmarkBold  & \color{red}\xmark    & \color{green}\CheckmarkBold & \color{green}\CheckmarkBold \\
Integer overflow/underflow (IF) \cite{IF}   &  \color{green}\CheckmarkBold  & \color{red}\xmark      & \color{green}\CheckmarkBold &\color{green}\CheckmarkBold \\
Arithmetic flaw (AF) \cite{AF}  &  \color{green}\CheckmarkBold  & \color{red}\xmark    & \color{green}\CheckmarkBold & \color{green} \CheckmarkBold  \\
Suicidal contract (SC) \cite{SC}  &  \color{green}\CheckmarkBold  & \color{red}\xmark  & \color{green}\CheckmarkBold &\color{green}\CheckmarkBold  \\
Ether leakage (EL) \cite{SC} &  \color{green}\CheckmarkBold  & \color{red}\xmark    & \color{green}\CheckmarkBold & \color{green}\CheckmarkBold \\ 
Insufficient gas (IG) \cite{IG}  &  \color{green}\CheckmarkBold  & \color{red}\xmark     & \color{green}\CheckmarkBold &\color{green}\CheckmarkBold  \\
Incorrect visibility/owner (IVO) \cite{IDV}  &  \color{green}\CheckmarkBold  & \color{red}\xmark  & \color{green}\CheckmarkBold &\color{green}\CheckmarkBold  \\\midrule
\multicolumn{4}{l}{\bf Functional Bugs\vspace{.1cm}}   \\
Price manipulation (PM) \cite{PM} & \color{green}\CheckmarkBold  & \color{green}\CheckmarkBold   & \color{red}\xmark & \color{green}\CheckmarkBold \\ 
Privilege escalation (PE) \cite{PE} & \color{green}\CheckmarkBold  & \color{green}\CheckmarkBold & \color{red}\xmark & \color{green}\CheckmarkBold \\    
Atomicity violation (AV) \cite{AV} & \color{green}\CheckmarkBold  & \color{green}\CheckmarkBold   & \color{red}\xmark & \color{green}\CheckmarkBold \\  
Business logic flaw (BLF) \cite{BLF} & \color{green}\CheckmarkBold  & \halfcirc & \color{red}\xmark & \color{green}\CheckmarkBold \\ 
Inconsistent state update (IS) \cite{DemystifyingBugs} & \color{green}\CheckmarkBold  & \color{green}\CheckmarkBold   & \color{red}\xmark & \color{green}\CheckmarkBold \\ 
Cross bridge (CB) \cite{CB}  & \color{green}\CheckmarkBold  & \halfcirc  & \color{red}\xmark & \color{green}\CheckmarkBold \\ 
ID uniqueness violation (IDV) \cite{DemystifyingBugs} & \color{green}\CheckmarkBold  & \color{green}\CheckmarkBold  & \color{red}\xmark & \color{green}\CheckmarkBold \\ 
\bottomrule

\end{tabular}
\end{table}
\section{Overview}
\label{sec:background_and_motivation}

This section first introduces the necessary background (\S\ref{subsec:background}), two motivating examples (\S\ref{subsec:motivating_examples}), followed by an overview of \sys's workflow (\S\ref{subsec:workflow}).

\subsection{Background}
\label{subsec:background}

This section introduces the modalities of smart contract, the definitions of implementation and functional bugs, smart contract invariants, and foundation models prompting. 

\vspace{.1cm}\noindent\textbf{Smart Contract Modalities.} Smart contract modality can be broadly understood as sources of information under which bug-preventive invariants are generated. Accordingly, smart contracts contain two main modalities: i) contract source code; ii) natural language, usually in the form of implementation-related comments and domain-specific textual information. From the contracts we studied in Table~\ref{table1}, 5,265 contracts (99.31\%) contain natural language related to code logic and expected transactional behavior. Existing invariants generators \cite{invariants} and bug detectors \cite{verismart, smartest, sailfish} focus only on a single modality, namely contract source code. To our best knowledge, \sys is the first smart contract analysis tool that can reason across both modalities.

\vspace{.1cm}\noindent\textbf{Bug Taxonomy.} As Table~\ref{bugtab} demonstrates, \sys-detected bugs can be categorized into two types: 

\begin{itemize}[leftmargin=*]
    \item \emph{Implementation bugs}: these bugs are generalizable by certain patterns in source code, such as reentrancy (RE), which can be generalized as a pattern of cyclic transactions. Reasoning about implementation bugs does not require domain-specific properties or multimodal information.
    
    \item \emph{Functional bugs}: theses bugs are tied to highly specialized transaction contexts and domain-specific properties. Detecting functional bugs requires understanding and reasoning across multimodal contract information. 
    Functions bugs are usually not pattern detectable from source code.
\end{itemize}

Implementation bugs usually do not require domain-specific information, so they can be detected based on general patterns~\cite{verismart, smartest, sfuzz, confuzzius, diffuzz, slither, mythril, manticore, echidna, bugsurvey, osiris, paymentbugs} without relying on any multimodal hints. 
For example, integer over/underflows (IF) can be detected by general test suites similar to buffer overflow \cite{bufferoverflow} used in traditional software. Reentrancy (RE) can be detected by testing general cyclic patterns of transactions. 

Functional bugs arise from highly domain-specific transactional contexts and exhibit unintended behavior under dynamic transaction executions, \ie incorrect stateful transitions and/or inter-contract communications given specific domains. Detecting functional bugs requires expert understanding of domain-specific properties and transaction contexts.
For example, a smart contract may contain some formula to calculate price, and a comment describing the transaction context is that ``price should stay within a certain range based on market trends from day 1 to day 30.'' 
These transaction contexts cannot be easily captured by code and are often overlooked by developers. 

\vspace{.1cm}\noindent\textbf{Smart Contract Invariants.} Invariants specify smart contract properties that should be held true throughout smart contract execution. Broadly, smart contract invariants can be categorized into two types, addressing implementation and functional bugs respectively. Invariants that track down implementation bugs can take the form of assertions specifying general patterns. Invariants that track down functional bugs often require templates tailored to domain-specific properties and transaction contexts. We have designed and built these invariant templates into \sys via a novel finetuning process and will detail them in \S\ref{subsec:multimodal_invariants_lib}.

\vspace{.1cm}\noindent\textbf{Prompting Foundation Models.} Foundation models (or Large Language Models) are pre-trained on texts and code with a large number of parameters. A foundation model seeks to estimate probabilistic distribution over tokenized data and generate new information based on seen data \cite{llm}. Besides the ``pre-train and finetune'' paradigm, ``pre-train, finetune, and prompt engineering'' paradigm shows high potential for challenging reasoning tasks\cite{prompt}. For example, given an invariant generation task, a prompt can be a question, such as ``what are the invariants at line 1?'' Recent research \cite{cot, cot2} has constructed prompts in different formats such that these prompts elicit intermediate reasoning steps of foundation models. The prompts used in \S\ref{subsec:ToT} elicit such intermediate reasoning steps.

\subsection{Motivating Examples}
\label{subsec:motivating_examples}
We provide two real-world hacks (simplified for readability) as motivating examples in this section.
We refer to the line number of a code statement as a \emph{program point} and the line number where invariants should be inserted as a \emph{critical program point}.  

\vspace{.1cm}\noindent\textbf{Example 1: Flashloan Primer.}  Flashloans in smart contracts are uncollateralized and allow users to borrow assets without any cost as long as users pay back \emph{within a single transaction}. Tokens declared with IERC library contracts and asset-swapping contracts that support external calls inherently support flashloans. The price manipulation (PM) bug in Listing~\ref{listing2} is confirmed by developers and multiple security firms \cite{visor, visor2, visor3}. Listing~\ref{listing2} demonstrates how malicious users take flashloans to exploit Visor, a money market contract providing liquidity services. Hackers first borrow a large \texttt{token0} flashloan, swap \texttt{token0} for \texttt{token1} on the platform to pump the price of \texttt{token1}, and as a result, inflate the price calculations at line 12. When \texttt{price} is sufficiently large, an attacker can mint and later withdraw a drastically inflated amount of shares by calling \texttt{deposit()}, where a user can mint shares by depositing a small amount of under-valued \texttt{token0} and a large amount of over-valued \texttt{token1}.  Although this hack is exacerbated by reentrancy \cite{wrongvisor1, wrongvisor2}, the root cause lies in flashloan based price manipulation.

For example, suppose the total balances of \texttt{token0} and \texttt{token1} are \$10 in \texttt{address(this)} with the token prices at \$1 each respectively. Without price manipulation, if Alice deposits \$10 \texttt{deposit0} and \$10 \texttt{deposit1}, she can mint \$20 shares, since the current \texttt{price} is \$1=10/10. To make a higher profit, Alice decides to take a flashloan of 1,000 \texttt{token0} (priced at \$1 per \texttt{token0}), swap flashloaned \texttt{token0} for \texttt{token1} in token pool reserves, resulting in inflated price of \texttt{token1} and deflated price of \texttt{token0}. Suppose the token prices of \texttt{token1} and \texttt{token0} are now \$100 and \$0.1 after the swap. When Alice invokes \texttt{deposit()} post swap, The \texttt{price} becomes \$1,000=(100*10)/(0.1*10), because the number of \texttt{token0} and \texttt{token1} in \texttt{address(this)} are still 10 but their individual token prices are manipulated. After the swap, Alice owns 1,000 \texttt{token1} (priced at \$100 per \texttt{token1}) and has 1,000 flashloaned \texttt{token0} (priced at \$0.1 per \texttt{token0}) as debt. Alice dumps all her \texttt{token1} as \texttt{deposit1} at an inflated sum of \$100,000 = 100*1000 and 10 \texttt{token0} as \texttt{deposit0} at \$1 = 10*0.1 via the \texttt{deposit()} call. Given the formula at line 17, this transaction allows Alice to mint shares totaling \$101,000=\$100,000 + 1*\$1,000. Alice pays back 1000 flashloaned under-valued \texttt{token0} (now only worth \$1,00 = 1,000* \$0.1), making a profit of \$100,900.
Listing~\ref{listing2} is a highly simplified example.  For interested readers, full mathematical details and developer's solution to the bug can be found at \cite{visormath, visorsol}.

Existing prompting frameworks \cite{gptscan, chatgpthowfar, manualaudit} specify no invariants and point to incorrect bugs such as reentrancy.
Existing bug analyzers based on formal verification, symbolic execution, and other dynamic analysis~\cite{verismart, smartest, smartest, invcon} report Listing~\ref{listing2} as a healthy contract, because they analyze only source code without considering the domain-specific price oracle context implied by the \colorbox{blueannoback}{blue comments}. 

 \mycode
\begin{lstlisting}[float,floatplacement=H,basicstyle=\fontsize{8}{9}\selectfont, caption={functional buggy snippet from the spot price manipulation of Visor  \cite{visor, visorethscan} (simplified for readability).}, label={listing2},numbers=left, xleftmargin=2em]
contract simplifiedVisor{
    /*two types of token reserves*/
    IERC20 token0, token1;
    /*reporting price at real time*/
    uint price;

    /*%*\colorbox{blueannoback}{real-time price updates}*) by the ratio of token reserves*/ 
    function getRealPrice() internal {
        //SmartInv: possible flashloan injection
        price = token1.balanceOf(address(this))/token0.balanceOf(address(this));
    }
    //SmartInv: minting shares by deposits
    function deposit(uint deposit0, uint deposit1, address to) public { 
        /*%*\colorbox{blueannoback}{price may change}*)*/
        getRealPrice(); 
        uint deposit0PricedInToken1 = deposit0 * price; 
        uint shares = deposit1 + deposit0PricedInToken1;
        if (deposit0 > 0) {
          token0.safeTransferFrom(msg.sender, address(this), deposit0);
        }
        if (deposit1 > 0) {
          token1.safeTransferFrom(msg.sender, address(this), deposit1);
        }
        ...
        _mint(to, shares);            
        }
     }
}

\end{lstlisting}
However, analyzing the Listing~\ref{listing2} bug requires understanding the natural language hints indicating that the price oracle is vulnerable to real-time price volatility.
Table~\ref{table:short_motivating_invariants} highlights \sys's solution using tailored invariant templates (discussion in \S\ref{sec:methodology}) by reasoning across source code and natural language hints.
\sys infers the lines immediately after lines 15 as critical program points, and infers an invariant \texttt{assert(price <= Old(price)*k)}. Similar to the use of \texttt{Orig()} in Daikon \cite{daikon} and \texttt{Old()} ESCJML \cite{ESCJML}, \texttt{Old(price)} returns the previous price point before the \texttt{deposit()} function is invoked. $k$ has a default value of 2 in \sys and can be updated based on developers' desired volatility ratio. Any violation of the assertion invariant would signal price volatility exceeding user desired $k$ and thus would signal price manipulation. 

One might argue that an attacker can manipulate the \texttt{Old(price)} first by taking flashloans and swapping tokens. However, our observation from thousands of blockchain transactions is that most users are honest. Therefore, before an attacker carries out such price manipulation attacks,  \texttt{Old(price)} generally returns a non-manipulated price point reflecting honest transactional activities at that moment. 

\begin{table}[!t]
\footnotesize
\setlength{\tabcolsep}{4pt}
\centering
\renewcommand{\arraystretch}{1.1}
\caption{\sys inferred invariants in Listing 2. 15+ refers to the line-numbered location where the invariant should be inserted, \eg 15+ means immediately after line 15. \texttt{Old(price)} evaluates a variable's pre-state and returns the price point before \texttt{liquidate} function is called. k is an adjustable ratio, where \sys sets default as k=2.}
\label{table:short_motivating_invariants}
\begin{tabular}{ll}
\toprule[1.1pt]
Critical Program Points & Inferred Invariants \\ \midrule[.9pt]
{15+}  & \texttt{assert(price <= Old(price)*k);}   \\
 \bottomrule[1.1pt]
\end{tabular}
\end{table}

\vspace{.1cm}\noindent\textbf{Example 2: Voting Fraud.} The voting fraud bug \cite{DemystifyingBugs, timelockcontroller} in Listing~\ref{example3} is officially recognized by the National Commmon Vulnerabilities and Exposures (CVE) with an assigned ID \cite{cve}. This hack was made possible by flashloans and classfied as priviledge escalation under functional bug types. The contract developers were aware of the potential for flashloan attacks, so they tried to mitigate the risk by restricting the order in which the \texttt{startExecute()}, \texttt{execute()}, and \texttt{endExecute()} functions could be invoked.

If a proposal is not ongoing and sTime = 0, then a message sender can invoke \texttt{startExecute()}.  If a proposal is ongoing (sTime != 0 and sTime + 24 hours $>$ block.timestamp), then a message sender can only invoke \texttt{execute()}. Otherwise (after 24 hours has passed), the proposal round can be ended by invoking \texttt{endExecute()}. As the contract developer(s) intended, \texttt{startExecute()} must be invoked before \texttt{execute()} within a proposal. \texttt{execute()} and \texttt{endExecute()} cannot be invoked within a single transaction (or within a 24-hour proposal round) to prevent flashloan attacks.

However, the key vulnerability lies in the developers' assumption that the three functions, \texttt{startExecute()}, \texttt{execute()}, and \texttt{endExecute()}, would be invoked sequentially in a proposal round. Unfortunately, that assumption does not hold. An attacker can bypass the \texttt{Execute()} function by invoking the \texttt{endExecute()} directly after taking a flashloan to become the highest proposer.

The attack above is possible, because \texttt{votingToken} variable is declared with the IERC20 wraparound library contract, which tracks how many tokens a user would like to vote when calling the \texttt{execute()} function. The IERC20 wraparound library contract also has its own \texttt{transferForm()} function. As a result, any variable declared with IERC20 can invoke \texttt{transferForm()} directly.

\begin{scriptsize}
\mycode
\begin{lstlisting}[float,floatplacement=H,basicstyle=\fontsize{7}{8}\selectfont, caption={openzepplin vulnerability (reported in CVE-2021-39168 and simplified for readability).}, label={example3},numbers=left, xleftmargin=2em]
contract TimelockController {
  /*this is a bidding contract: 
  watch out for %*\colorbox{blueannoback}{flashloan}*)*/ 
  struct Proposal {
    uint sTime; address newOwner;
  } 
  IERC20 %*\colorbox{blueannoback}{votingToken;}*) /*important variable*/
  address owner;
  Proposal proposal;
  
  /*the following three functions should be %*\colorbox{blueannoback}{executed atomically}*)*/
  function startExecute() external {
    require(proposal.sTime == 0, "on-going proposal");
    proposal = Proposal(block.timestamp, msg.sender);
  }
    
  function execute(uint amount) external {
    require(proposal.sTime + 24 hours > block.timestamp, "execution has ended");
    votingToken.transferFrom(msg.sender, address(this), amount); 
  }
  
  function endExecute() external {
    require(proposal.sTime != 0, "no proposal");
    require(proposal.sTime + 24 hours < block.timestamp, "execution has not ended");
    require(votingToken.balanceOf(address(this)) * 2 > votingToken.totalSupply(),"execution failed");
    /*we're about to %*\colorbox{blueannoback}{change the owner}*) of the contract */ 
    owner = findHighest(_allProposals);
    delete proposal;
  }   
    
  /*highest proposer becomes the new owner of the contract and gets all locked funds*/ 
  function getFunds() external onlyOwner {
    ...
    return allLockedTokens; 
  }
}
\end{lstlisting}
\end{scriptsize}
Suppose a hacker borrows a large flashloan and injecting it into \texttt{votingToken} via the \texttt{transferForm()} function to make the highest bid. This bypasses the \texttt{execute()} function and allows the hacker to invoke \texttt{transferForm()} directly in the IERC20 library contract. After 24 hours, the hacker ends the proposal round by invoking the \texttt{endExecute()} function. This exploit allows the hacker to become the new owner of the contract at line 27, and thus to invoke the highly privileged \texttt{getFunds()} function at line 32. 
Then the hacker can obtain all locked tokens and pay back the flashloan with a profit.

Existing analyzers \cite{verismart, smartest, manticore} mistakenly report that Listing~\ref{example3} contract contains an integer overflow/underflow bug related to $sTime$ at lines 23 and 24 (false positives because Solididy version $\geqslant$ 0.8 automatically preempts operations causing integer overflow/underflow) while omitting the more damaging privilege escalation bug. Their mistaken reporting stems from relying on pattern-matching arithmetic operations without considering the underlying transactional logic.


\sys's solution is to reason across source code and natural language hints in \colorbox{blueannoback}{blue}(comments and variable names related to the transactional context). First, from the source code and comments, \sys infers that the transactional context is ``bidding.'' After predicting "bidding" transactional context, \sys infers critical program points and invariants as highlighted in Table~\ref{table:motivating_invariants}. If a malicious actor bypasses the \texttt{execute()} function and injects a large flashloan in \texttt{endExecute()} function directly, then \texttt{votingToken} variable would transition to a wrong state that violates the assertion invariant (\texttt{Old(votingToken.balanceOf(address(this))}
 \texttt{==votingToken.balanceOf(address(this)}) after line 25.

\subsection{\sys Workflow} 
\label{subsec:workflow}
\begin{figure}[!t]

\includegraphics[width=\linewidth]{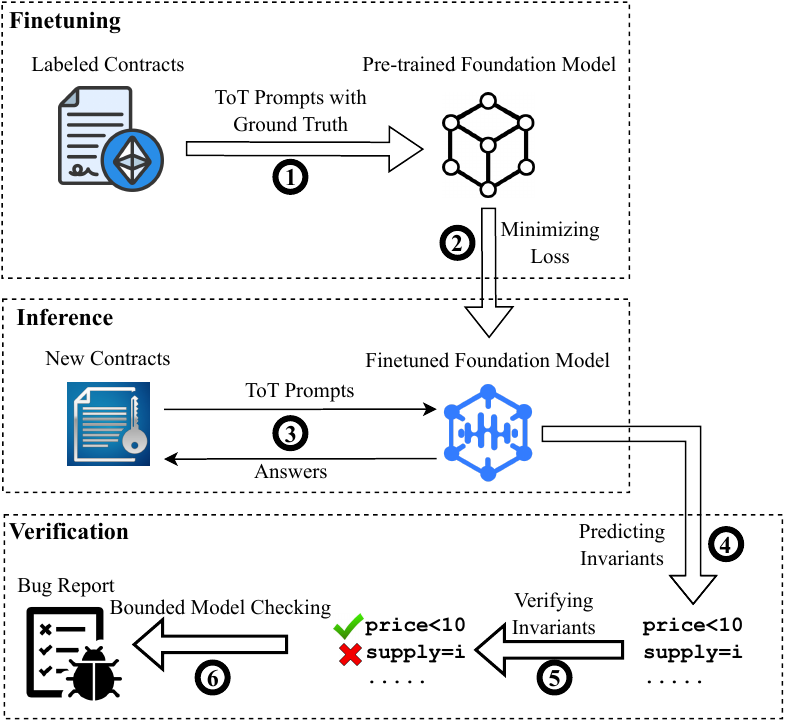}

\caption{\sys's Workflow}
\label{fig:design}
\end{figure}

Figure~\ref{fig:design} shows \sys's workflow. 
\sys first finetunes the model on a dataset of labeled contracts with Tier of Thought (ToT) prompts and ground truth at \textcircled{\footnotesize{1}}. 
\begin{table}[!t]
\footnotesize
\setlength{\tabcolsep}{1.5pt}
\centering
\renewcommand{\arraystretch}{1.1}
\caption{\sys inferred invariants in Listing 3. 19+ and 25+ refer to the line-numbered location where the invariant should be inserted, \eg 19+ means immediately after line 19. \texttt{Old(votingToken.balanceOf(address(this)))} returns the votingToken balance in \texttt{address(this)} before the instrumented function is called. At 19+, inferred invariant specifies that the current balance of \texttt{votingToken} equals to the sum of transferred amount and the prior balance before the transfer. At 25+, the inferred invariant specifies that total balance of \texttt{votingToken} stays the same after a proposal round has ended, \ie no flashloan transfers into \texttt{votingToken}.}
\label{table:motivating_invariants}
\scalebox{0.9}{
\begin{tabular}{ll}
\toprule[1.1pt]
Critical  & Inferred  Invariants\\ 
Program Points & \\ \midrule[.9pt]
{19+}  & \texttt{assert(votingToken.balanceOf(}  \\ 
 & \texttt{address(this)))==}  \\ 
 & \texttt{Old(votingToken.balanceOf(}  \\ 
 & \texttt{address(this)))+amount);}  \\ 
\midrule[.9pt]
{25+}  & \texttt{assert(Old(votingToken.balanceOf(}   \\
    & \texttt{address(this)))==votingToken.balanceOf(} \\
    & \texttt{address(this)));} \\
 \bottomrule[1.1pt]
\end{tabular}}
\end{table}
\sys learns to minimize cross entropy loss \cite{cel}
between ground truth and inferred answers at \textcircled{\footnotesize{2}}.
During inference, \sys takes a previously unseen new contract as input and prompts the finetuned model using ToT at \textcircled{\footnotesize{3}}. 
We develop a new iterative prompting process: \sys uses the answers from prior easier tiers to guide answer generation for subsequent more challenging tiers.
After the finetuned model generates invariants at \textcircled{\footnotesize{4}}, \sys proceeds to verify inferred invariants by proving program correctness at \textcircled{\footnotesize{5}} first.
If no proof of program correctness is found after the initial verification, \sys uses a bounded model checker to seek violations (counterexamples) of inferred invariants at \textcircled{\footnotesize{6}}. As a final step, \sys outputs a report on verified invariants and detected bugs. 
Once finetuned, \sys is fully automated to detect bugs.

In building this workflow, there are two technical challenges. 
The first one is \emph{how to incorporate and represent multimodal information} that also respects smart contract semantics during finetuning. 
Our evaluation shows that simply prompt engineering without customized training datasets cannot identify correct invariants in real-world contracts. 

To overcome the challenge, we have designed tailored invariant templates (in \S\ref{subsec:multimodal_invariants_lib}) and built a unique finetuning process (in \S\ref{subsec:ToT}) that incorporates multimodal information. Our finetuning process tailors answers to ToT-prompts. 
Furthermore, foundation models are known to have the hallucination problem \cite{hallucination}. Thus, a second challenge is \emph{to determine which invariants are correct} during inference on previously unseen contracts without labeled ground truth. To overcome the second challenge, \sys adopts novel invariants ranking strategy for effective verification (in \S\ref{subsec:verification_algorithm}).  
\vspace{.1cm}\noindent\textbf{Finetuning and Ground Truth.} Finetuning models to both consistently reason about diverse sets of invariants and to detect a wide range of real-world bugs is non-trivial. Given that no prior foundation-model-based work has done both (to our best knowledge), SmartInv presents the first multimodal-reasoning-based finetuning approach for invariants generation and bug detection. We specifically focus on modalities helpful to detect functional bugs unique to smart contracts.

\sys uses ToT's step-by-step reasoning and tailored ground truth to finetune the pre-trained model. Each training sample consists of a smart contract file collected from Etherscan \cite{etherscan} and annotated ground truth of six labeled features (details in Table~\ref{table:ground truth}). 
During finetuning, an input contract, each ToT prompt, and corresponding ground truth are encoded as token sequences. 
For example, a ToT prompt with ground truth answers can be ``What are the critical program points in the contract? Critical program points are [ground truth label].'' 
\sys splits the training dataset into train, validation, and test sets of 1381, 296, 296 contracts (70\%, 15\%, 15\%) respectively.

\vspace{.1cm}\noindent\textbf{Inference.} The finetuning design above facilitates \sys's unique inference approach: using iterative and increasingly complex prompts that respect smart contracts' domain-specific properties and semantics, \eg transactional context and critical program points. During inference, the finetuned model is prompted with a previously unseen contract and a tiered prompt without any answers, such as ``What are the critical program points in the contract?'' The finetuned model can predict critical program points as an answer, because the model is finetuned for this specific downstream task. Then \sys uses inferred answers from prior prompts to elicit answers on  more challenging prompts. 

\vspace{.1cm}\noindent\textbf{Verifying Predicted Invariants.} After inference, \sys uses a new invariants ranking strategy for effective verification: it prompts the model to rank inferred invariants from most likely to be correct and bug preventive to the least likely. After ranking, \sys automatically switches to a verifier that tries to prove for program correctness on ranked invariants. If such proofs can be found on an inferred invariant, \sys marks that invariant as correct. Otherwise, \sys uses bounded model checker to search for violations of inferred invariants. When violations are found, such violations signify two scenarios that warrant further review: i) a potential bug; or ii) potentially incorrect invariants. Therefore, we inspect the counterexamples to i) confirm the existence of bugs and thereby correct invariants or ii) confirm the incorrectness of inferred invariants. 
If no correctness proofs or counterexamples are found, \sys discards that unproven invariant.



\section{Methodology}
\label{sec:methodology}
We formally define the invariant inference problem in \S\ref{subsec:problem_formulation}. \sys is finetuned with customized invariant types in \S\ref{subsec:multimodal_invariants_lib} to facilitate the unique Tier of Thought finetuning process in \S\ref{subsec:ToT}. As a result, \sys learns to use these invariant types to answer prompted questions and to infer invariants in an iterative fashion during inference. With novel invariants ranking strategy, \sys uses the  verification algorithm in \S\ref{subsec:verification_algorithm} to verify inferred invariants. 

\subsection{Problem Formulation}
\label{subsec:problem_formulation}



Let $M_\theta$ be the pre-trained foundation model parameterized by $\theta$ and let $S$ be tokenized input contract. 
Let $C$ be tokenized program points (\ie a line-numbered location of a code statement), $V$ be invariants and instrumentations for invariant checks respectively. For simplicity, we refer to invariants and related instrumentations as invariants in this section. From input $S$, 
we finetune (train) the model $M_{\theta}$ to generate critical program points $c_i$ and associated invariants $v_i$ as ($c_i$, $v_i$), where $c_i \in C$ and $v_i \in V$. From predicted ($c_i$, $v_i$), we further finetune $M_{\theta}$ to predict vulnerabilities in $S$.  

\begin{scriptsize}
\mycode
\begin{lstlisting}[float,floatplacement=H,basicstyle=\fontsize{7}{8}\selectfont, caption= training example contract, label={train_example},numbers=left, xleftmargin=2em]
contract trainingExample {
 /*%*\colorbox{blueannoback}{totalSupply and balances should be same}*) before and after transfer*/
 uint totalSupply, tokens;
 %*\colorbox{blueannoback}{mapping(address$\implies$uint)} *)balances;

 function transfer(address to) external {
  balances[to]=%*\colorbox{blueannoback}{balances[to].add(tokens)}*);  
  balances[msg.sender]=%*\colorbox{blueannoback}{balances[msg.sender].sub(tokens)}*);   
  }
  
 /*%*\colorbox{blueannoback}{only contract owner}*) should invoke tokenIncrease*/ 
 function tokenIncrease() external {
  if (tokens<=100) {
    tokens+=1.1*tokens;
   }
  return tokens;
 }
 ...
}
    
\end{lstlisting}
\end{scriptsize}
\label{subsec:ToT}
\subsection{Invariant Types}
\label{subsec:multimodal_invariants_lib}
Broadly, \sys infers three types of invariants  to capture functional bugs: assertions with special expressions, modifiers, and global invariants. These invariants are highly generalizable and can be easily verified by the state-of-the-art verifiers\cite{verx, cryticinv, verisol}. Listing~\ref{train_example} and Table~\ref{table:ground truth} illustrate the use of a subset of  these invariant types during finetuning. 

A common invariant type inferred by \sys is \texttt{assert(expr1 op expr2)} at a critical program point. \texttt{expr1} and  \texttt{expr2} are legal Solidity expressions. \texttt{op} are binary operators, such as ==, $>$=, $<$=, !=. Assertions can also be replaced by pre-condition check \texttt{Assume(expr)}, as well as post-condition checks \texttt{Ensures(expr)} and \texttt{require(expr)}, because \sys's backend verifier also supports these additional checks. As part of assertions, \sys also infers special expressions uniquely tailored to smart contracts, such as \texttt{Old(expr)}, \texttt{k*Old(expr)}, and \texttt{SumMapping(mappingVar)}.

The use of \texttt{Old(expr)} is similar to Daikon's \texttt{Orig(expr)}\cite{daikon}. It returns the previous state of a variable at the entry of a function. Ratio \texttt{k} allows users to specify an accepted volatility ratio for a variable, usually a price point. \sys sets the default volatility ratio $k$ to 2. Finetuning \sys to learn \texttt{Old(expr)} and \texttt{k} enables invariant inference related to inconsistent state updates and price manipulation bugs. To check arithmetic operations across mapping and integer/byte types, \sys sums up the values stored in multi-layered maps using customized \texttt{SumMapping(mappingVar)}. This design enables \sys to directly compare mapping type against other primitive types. 

Modifiers are invariant-like Solidity functions that specify the behavior of other functions. \sys infers function modifiers, because they are useful to express expected behavior of an entire function beyond assertions. Take Table~\ref{table:ground truth} as an example. \sys infers an \texttt{onlyOwner} modifier at critical program point 10+. When the \texttt{tokenIncrease()} unction at line 12 is instrumented with the modifier, only contract owner can invoke \texttt{tokenIncrease()} function. Intuitively, modifiers are function-level invariants.

To specify cross-function and cross-contract behavior, \sys also learn \texttt{Invariant(expr)}, a customized invariant function that specifies expected return values of a function during cross-contract calls. \eg \texttt{Invariant(func()==a)}. Compared to assertions that can only specify program behavior at a given program point, \texttt{Invariant(expr)} can also specify loop invariant when placed at the beginning of loops and specify state variables outside functions. \texttt{Invariant(expr)} is thus versatile.

\begin{table}[!t]
\footnotesize
\setlength{\tabcolsep}{1.2pt}
\centering
\renewcommand{\arraystretch}{1.1}
\caption{Labeled ground truth for the trainingExample contract. Repeated code fragments are replaced by ... in ``Critical Invariants'' and ``Ranked Critical Invariants'' labels. ``Rank 1'',``Rank 2'', and ``Rank3'' refer to a group of invariants that can discover bugs in descending likelihood.}
\label{table:ground truth}
\scalebox{0.9}{
\begin{tabular}{ll}
\toprule[.6pt]
\textbf{Labeled Features} & \textbf{Ground Truth} \\ 
\midrule[.6pt]
transactional context  & token transfer  \\ 
\midrule[.6pt]
critical program points & 7+, 8+, 10+, 12, 17+ \\
\midrule[.6pt]
Invariants
    &7+ \texttt{assert(balances[msg.sender]>=tokens);}\\

    &8+ \texttt{assert(sumMapping(balances)==}\\
    & \hspace{0.5cm}\texttt{totalSupply);} \\
    &10+ \texttt{modifier onlyOwner\{}\\  &\hspace{0.5cm}\texttt{require(msg.sender==owner);\}; } \\
    &12 \texttt{function tokenIncrease()}\\
    & \hspace{0.5cm}\texttt{\emph{onlyOwner} external \{...\};} \\
     &17+ \texttt{Invariant(tokenIncrease()>100);}\\
\midrule[.6pt]
Critical 
Invariants &7+ \texttt{assert(...);}\\
    &8+ \texttt{assert(...);}\\
    &10+ \texttt{modifier onlyOwner\{...\}; } \\
    &12 \texttt{function tokenIncrease(uint tokens)}\\
    &\hspace{0.5cm}\texttt{\emph{onlyOwner} external \{...\}; } \\
\midrule[.6pt]
Ranked  & Rank 1: 10+ \texttt{modifier onlyOwner\{...\}};  \\
Critical &12 \texttt{function tokenIncrease()}\\
Invariants &\hspace{0.5cm}\texttt{\emph{onlyOwner} external \{...\}; } \\
    &Rank 1: 7+ \texttt{assert(...);}\\
    &Rank 2: 8+ \texttt{assert(...);} \\
    &Rank 3: 17+ \texttt{Invariant(...);} \\
\midrule[.6pt]
Vulnerabilities  & incorrect visibility/ownership; arithmetic flaw; \\

\bottomrule[.6pt]
\end{tabular}}
\end{table}

\subsection{Tier of Thought Finetuning and Inference}

To guide a pre-trained foundation model towards generating bug-preventive invariants, the key innovation is to introduce increasingly complex thoughts to reason from contract source files to correct answers of each prompt. Given input contract, each thought is a tokenized sequence, such as ``What is the transactional context in the contract? The transactional contract is token transfer.'' 

\vspace{.1cm}\noindent\textbf{Finetuning with ToT.} We finetune the model to generate one thought at a time, starting with the simplest and working our way up to the most complex thoughts. Taking Listing~\ref{train_example} and Table~\ref{table:ground truth} as an illustrative training example for this section, the model learns to reason about smart contracts' source code and natural language highlighted in \colorbox{blueannoback}{blue} (variable names and comments useful for program understanding and invariants generation). Using multimodal information, the model is finetuned to generate tokenized answers given a prompt. This design enables \sys to predict domain-specific information (\ie the ground truth of labeled features) on new contracts. 
\sys adds ``$<$end of text$>$" as a special token to separate each training sample. 

\vspace{.1cm}\noindent\textbf{Tier 1 Finetuning (Critical Program Points)}.
In this tier, the model tokenizes contract source files, tier 1 prompts, and answers from labeled ground truth as sequences. The tier 1 thoughts below seek to finetune \sys's understanding of transactional contexts and critical program points from multimodal sources in the contract. The example below illustrates tier 1 training sample: 

\vspace{0.1cm}\noindent\fcolorbox{black}{gray!20}{\begin{minipage}{24em}
\texttt{Contract trainingExample \{...\}}\\
What's the transactional context of the contract? The transactional context is \textbf{token transfer}.\\
Given transactional context, what are the critical program points? Critical program points are \textbf{7+, 8+, 10+, 12, 17+}.\\
$<$End of Text$>$
\end{minipage}}\vspace{0.1cm}

\vspace{.1cm}\noindent\textbf{Tier 2 Finetuning  (Invariants)}. At tier 2, the model tokenizes contract source files, tier 2 prompts, and the ground truth of ``Invariants'' and ``Critical Invariants'' labels for finetuning. Critical invariants refer to those that are likely to prevent bugs. \sys's tier 2 finetuning design continues the thoughts from tier 1 and facilitates invariants generation at correct program points during inference.  The example in grey box illustrates the training sample design of tier 2:

\vspace{0.1cm}\noindent\fcolorbox{black}{gray!20}{\begin{minipage}{24em}
\texttt{Contract trainingExample \{...\}}\\
Given inferred critical program points, what are the invariants? Invariants are [ground truth of  ``Invariants'' label in Table~\ref{table:ground truth}].\\
Given inferred invariants, what are the critical invariants? Critical invariants are 
[ground truth of  ``Critical Invariants'' label in Table~\ref{table:ground truth}].\\
$<$End of Text$>$
\end{minipage}}\vspace{0.2cm}

\texttt{assert(sumMapping(balances)==
totalSupply)} at critical program point 8+ in Table~\ref{table:ground truth} is derived from both natural language cues and source code.
This invariant checks that the condition specified in natural language at line 2 holds true. \sys uses a special expression \texttt{sumMapping(*)} to sum up all balances stored in the \texttt{balances} mapping. This way, \sys can directly compare \texttt{balances} of mapping type with  \texttt{totalSupply} of uint type. The second natural language inspired invariant is \texttt{modifier onlyOwner(...)} at critical program point 7+, with added \texttt{onlyOwner} modifier instrumentation to the \texttt{tokenIncrease} function at line 12. This pair checks that only the contract owner can invoke \texttt{tokenIncrease} function as hinted by comments at line 11. By these two invariant pairs, \sys learns natural language hints and associated invariants during finetuning. 

The remaining ground truth of the ``Invariants'' label in Table~\ref{table:ground truth} are based on source code. The invariant \texttt{assert(balances[msg.sender])>=
tokens)} at program point 7+ checks that a message sender has enough balance to make the transfer. \texttt{Invariant(tokenIncrease()>100)} at program point 17+ checks that the return value of \texttt{tokenIncrease} function is great than 100 during cross-contract calls. 

After being finetuned to generate invariants, \sys is also finetuned to generate critical invariants that can potentially identify bugs. The invariant \texttt{assert(balances[msg.sender])>=
tokens)} checks against arithmetic flaws. The invariant \texttt{modifier onlyOnwer\{...\}} with function signature instrumentation checks against incorrect access control. Thus they are labeled as critical invariants for Listing~\ref{train_example} training contract. By comparison, the invariant at 17+ \texttt{Invariant(tokenIncrease()>100)} checks function return value without likely bugs in this sample. Therefore, 17+ \texttt{Invariant(...)}  is not included in the ground truth of ``Critical Invariant'' label. With critical invariants, \sys learns to reason about likely invariants that can guard vulnerable code fragments effectively.  

\vspace{.1cm}\noindent\textbf{Tier 3 Finetuning (Prioritized Invariants and Bugs)}. \sys is also finetuned to rank/prioritize critical invariants and predict vulnerabilities in the contract from previously generated information. The example below illustrates the construction of tier 3 training sample:

\vspace{0.1cm}\noindent\fcolorbox{black}{gray!20}{\begin{minipage}{24em}
\texttt{Contract trainingExample \{...\}}\\
What are the ranks of inferred critical invaraints? The ranks of inferred critical invariants are [ground truth of ''Ranked Critical Invariants'' label in Table~\ref{table:ground truth}].\\
What are the vulnerabilities in the contract? The vulnerabilities are [ground truth of ``vulnerabilities'' label in Table~\ref{table:ground truth}].\\
$<$End of Text$>$
\end{minipage}}\vspace{0.2cm}

Specifically for Listing~\ref{train_example}, the first rank 1 invariants at critical program points 10+ and 12 identify an incorrect visibility/ownership bug. The second rank 1 invariant at critical program points 7+ identifies an arithmetic flaw bug: the contract lacks proper guard to ensure that a message sender has sufficient balances to make a token transfer. Invariants of rank 2 and rank 3 are correct but trivial invariants that are less likely to find bugs.

\vspace{.1cm}\noindent\textbf{Inference with ToT.} The unique aspect of \sys's inference is to decompose the invariant inference problem into three-tiered tasks, thereby making an iterative process. \sys uses inferred answers from the previous tier to generate answers for more challenging prompts of later tiers. At each tier, \sys prompts the finetuned model on a previously unseen contract with two tailored prompts.

At tier 1, \sys tackles the simple task by inferring transactional contexts and critical program points first. \sys uses the answer generated for prompt A to continue generating critical program points for prompt B.

\vspace{0.1cm}\noindent\fcolorbox{black}{gray!20}{\begin{minipage}{24em}
\textbf{Tier 1 Prompts}\\
Prompt A: What's the transactional context of the contract?\\
Prompt B: Given transactional context, what are the critical program points?
\end{minipage}}\vspace{0.1cm}

At tier 2, \sys is given a slightly more complex task of inferring invariants at predicted critical program points and identifying critical invariants, \ie bug-preventive invariants from all inferred invariants.

\vspace{0.1cm}\noindent\fcolorbox{black}{gray!20}{\begin{minipage}{24em}
\textbf{Tier 2 Prompts}\\
Prompt A: Given inferred critical program points, what are the invariants?\\
Prompt B: Given inferred invariants, what are the critical invariants? 
\end{minipage}}\vspace{0.2cm}

At tier 3, \sys is given the most challenging task: ranking critical invariants for verification and predicting vulnerabilities in a contract. \sys reports verified invariants with predicted vulnerabilities as a final report. We highlight that buggy traces from a verifier is more sound than model inferred bugs. However, because verifiers frequently encounter incompatible Solidity compilers, \sys reports predicted vulnerabilities as a remedy.  

\vspace{0.1cm}\noindent\fcolorbox{black}{gray!20}{\begin{minipage}{24em}
\textbf{Tier 3 Prompts}\\
Prompt A: What are the ranks of inferred critical invaraints?\\
Prompt B: What are the vulnerabilities in the contract? 
\end{minipage}}\vspace{0.2cm}

\subsection{ToT Invariants Verification Algorithm}
\label{subsec:verification_algorithm}
\begin{algorithm}[!t]
\footnotesize
\setlength{\tabcolsep}{2pt}
	\caption{ToT Invariants Verification}
\label{algo: verify}
\begin{tabular}{lp{2.9in}}
\textbf{Input}:
    & an input smart contract $S$, Tier 3 invariants ranking model $M_{\theta}$, and an initial set of assumed true positive candidate invariants $V$ generated by $M_{\theta}$. \\
\textbf{Output}:
    & verification report $P$,  a set of verified correct invariants $I_{correct}$, and a set of possibly correct invariants that require further inspection  $I_{possible}$.
\end{tabular}

\begin{spacing}{1.1}
\begin{algorithmic}[1]
\State $Syntax\_Check(V);$
\While{$V \neq \emptyset$}
\State  $v_i \gets$  Tier3.rank(); \color{blue}\Comment{ranking critical invariants}\color{black}
\State  $V \gets V - {v_i}$;
\State  $ \Pi(\sigma(v_i), \mu)$ = $Inductive\_Check (S, v_i)$;
      \If {$\sigma(v_i)==True$}  \color{blue}\Comment{proof of program correctness found}\color{black}
\State    $I_{correct}.append(v_i)$;
\State       $P = P.append(\Pi)$\;
     \Else \color{blue}\Comment{proof of program correctness not found}\color{black}
\State       $\Pi(\phi, \overline{\mu})$ = $BMC(S, v_i, m)$; \color{blue}\Comment{entering BMC phase}\color{black}
\State       $P = P.append(\Pi)$;
      \If {$\overline{\mu}$ is a counter example} 
\State             $I_{possible}.append(v_i)$;
\State              break;  \color{blue}\Comment{manual inspection requested for counterexamples}\color{black}
      \EndIf

\State      discard $v_i$; \color{blue}\Comment{no counterexamples or correctness proof found}\color{black}
 \EndIf   
\EndWhile
\State return $P$, $I_{correct}$, and $I_{possible}$;

\end{algorithmic}
\end{spacing}
\end{algorithm}

Algorithm~\ref{algo: verify} has three phases to verify inferred invariants from ToT: candidate invariants ranking; proving program correctness by induction; bounded model checking. $S$ denotes an input contract; $V$ denotes a set of candidate invariants; $v_i$ denotes a tier-3 ranked critical invariant selected from $V$; $I_{correct}$ denotes verified correct invariants; $I_{possible}$ denotes possibly correct invariants that require human inspection. From a set of candidate invariants $V$ and an input contract $S$, the algorithm discards the invariants that cause compilation errors at line 1 first. Then tier 3 ranks critical invariants from candidate invariants at line 3. The novelty of Algorithm~\ref{algo: verify} is using foundation model guidance (critical invariants ranking) for verification, and as a result, increases verification efficacy.

\vspace{.1cm}\noindent\textbf{Induction Phase.} After \sys unrolls ranked invariants at line 3 using tier 3's prompt A, the $Inductive\_Check()$ function at line 5 maps Solidity to Boogie and uses Boogie's monomial predicate abstraction \cite{boogie, Lahiri} to check whether $v_i$ is inductively strong enough to prove program correctness. \sys selects Boogie for two specific features: i) it has built-in map types, e.g., [int] bool. These map types correspond well to Solidity’s mapping types, e.g., mapping(int=$>$bool). Smart contracts use mappings frequently. ii) Boogie produces precise failing traces for each buggy procedure. This enables easy tracking on what kind of input triggers invariant violations.

If an invariant is inductively strong enough, Boogie will generate a proof of correctness $\mu$. In this case, \sys adds $v_i$ to  $I_{correct}$ as verified correct invariants and the full verification result $\Pi$ is added to the final report $P$. If an invariant is not inductively strong enough, the bounded model checker will search for counterexamples.

\vspace{.1cm}\noindent\textbf{Bounded Model Checking Phase.} If invariant $v_i$ cannot be proven inductive, \ie $\sigma(v_i)$ is false, our bounded model checker $BMC(...)$ leverages \corral\xspace to search for counterexamples $\overline\mu$. If counterexamples are found on $v_i$, there are two possible cases: i) $v_i$ is a correct (true positive) invariant and a bug is found by the merit of counterexamples; or ii) $v_i$ is an incorrect invariant. Therefore, we break the loop for further inspection. This algorithm is applied iteratively until all ranked critical invariants are evaluated. 

During verification, \sys produced 51,505 counterexamples on 89,621 real-world contracts. Upon review, 46,360 (90.01\%) counterexamples resulted from confirmed bugs and 5,145 (9.99\%) counterexamples were due to incorrectly inferred invariants.

\section{Implementation}
\label{sec:implementation}

We implemented \sys's training and inference in 4011 lines of Python, the invariant verification algorithm in 1322 lines on top of \verisol.

\vspace{.1cm}\noindent\textbf{Model Optimization and Hardware.} We selected LLaMA-7B \cite{LLaMA} as the backbone of \sys. To enable memory-efficient training, we applied 8-bit quantization \cite{quantization}, Parameter Efficient Finetuning (PEFT) \cite{peft}, and low-rank adaptation (LoRA) \cite{lora} to LLaMA-7B during finetuning. This optimization allowed our model to complete training within a single Nvidia RTX 2080Ti GPU, as opposed to usual requirements of 4 A6000 GPUs. 

We ran all experiments and evaluations on a Linux Server with Intel Xeon 4214 at 2.20GHz with 48 virtual cores, 188GB RAM, and 4 Nvidia RTX 2080Ti GPUs, a Google coLab plus account with additional computation units, and a commercial server with 4 A6000 GPUs.

\vspace{.1cm}\noindent\textbf{Dataset.} We collected source files of 179,319 contracts in total, covering a period from January 1, 2016 to July 1, 2023. Of those 179,319 contracts, 175,991 contracts were crawled from Etherscan \cite{etherscan} via Google BigQuery and 3,328 contracts were crawled from 78 live decentralized applications (dApps)' public git repositories. We selected 572 contracts (2,173 annotated samples post ToT data augmentation) that represented each bug type in Table~\ref{bugtab} for training. 

For evaluation, we excluded 89,698 contracts that are: i) duplicates; ii) require old Solidity compilers ($<$ 0.3.x); iii) written in non-Solidity languages (Vyper and Go); iv) already included in our training dataset. Thus, our evaluation dataset consists of 89,621 Solidity contracts averaging 1,621 lines of code per contract and they are different from our training dataset. Following \cite{sailfish}, we categorized our evaluation dataset into three subsets based on lines of code: i) \emph{small}: [0, 500); ii) \emph{medium}: [500, 1000); iii) \emph{large}: [1000, $\infty$), consisting of 65,739, 12,011, and 11,871 contracts respectively. 

\vspace{.1cm}\noindent\textbf{Labeled Features.} To provide domain-specific insights, we labeled the ground truth of each training contract with six features: transactional contexts, critical program points; all relevant invariants; critical invariants, ranked critical invariants; vulnerabilities if the contract contains any. 
As illustrated in \S\ref{sec:methodology}, we embedded the ground truth via ToT prompts during training. 
Specifically, we have labeled contracts with the following ten transactional contexts in our training dataset: ERC libraries, token transfer, cross bridge, bidding, voting, lottery, healthcare, investing, price oracle, and other. These labels cover the top use cases as identified by Ethereum \cite{usecase}. 
To ensure correct critical program points and invariant labels, we ran the verification algorithm in \S\ref{subsec:verification_algorithm} and cross-checking with at least two researchers. To ensure correct vulnerabilities labels, we reproduced 3213 hacks in 4033 lines of Solidity on a forked ethereum virtual machine (EVM) \cite{defihacklabs} to confirm the existence of labeled vulnerabilities.

\vspace{.1cm}\noindent\textbf{Hyperparameters.} For model optimization, we set LoRA\_alpha = 32, lora\_dropout = 0.01, LoRA\_R = 8, learning rate = 3e-4, micro\_batch = 1. During inference, we set temperature, top-k, top-p, and repeated penalty to 0. The hyperparameters of other three finetuned foundation models are documented in our github README.

We also note that pre-trained LLaMA cannot process contracts beyond 20 lines due to limited token length. Finetuning breaks such limitation by adding additional token mappings from the initial 512 to 4096. Therefore, our finetuned LLaMA can reason about medium ([500, 1000) lines) contracts. On large ([1000, $\infty$) lines) contracts, finetuned LLaMA are still limited by available token length. In that case, we prompted the model to summarize imported modules, \ie imported library and helper contracts, to fit in available tokens.

\section{Evaluation}
We evaluate \sys to answer following questions: 
\begin{itemize}[leftmargin=*]
\item\textbf{RQ1}: In terms of bug detection, how does \sys compare to six prior bug analyzers and three similar prompting-based tools?

\item\textbf{RQ2}: In terms of invariants generation, how does \sys compare to similar tools? 

\item\textbf{RQ3}: How much do our selected model LLaMA and optimizing strategies improve the accuracy of bug detection and invariants generation? 

\item\textbf{RQ4}: How fast is \sys compared to similar tools?
\end{itemize}

\vspace{.1cm}\noindent\textbf{Experiments Setup.} To sufficiently represent available tools, we selected bug analyzers covering a wide range of techniques with minimal overlapping.
We installed and followed the instructions of the latest versions (as of July 28, 2023) of each bug analyzer from their git repositories.

 \vspace{.1cm}\noindent\textbf{Ground Truth Measurement.} We define ground truths in two relevant aspects: bugs and invariants. For bug detection, we conducted both large-scale and refined experiments. The large-scale experiment summarized each tools' reported results. We further validated the reported results by manually reviewing a subset of projects in the refined experiment. For invariants generation, we inspected the invariants generated in the refined experiments to gain a granular understanding. The scale of our analysis and ground truth measurement are in line with previous work \cite{sailfish, verismart, smartest}.

Each tool under evaluation scanned the 89,621 contracts in our large-scale experiment and we recorded their reported results in Table~\ref{table:reported_bugs}. We acknowledge that results in Table~\ref{table:reported_bugs} summarize each tool's reported bugs, which may not necessarily be true positive or exploitable. To gain more insights into each tool's false positive/negative ratio, we conducted refined experiment on 60 well-known hacked projects (1,241 buggy contracts) and used their audit reports as ground truths for bugs. We recorded correct (TP), incorrect (FP), and missed (FN) alarms on bugs in the contract. To determine ground truth of detect bugs by each model, we defined ``Accuracy (Acc.)'' on a per-contract basis: we marked an output as accurate only when a model generated bug-preventive invariants at correct program points for an entire contract and inferred the correct bugs.

We reported three outcomes on each contract and grouped the results by bug type: i) \emph{bugs}: the number of a given bug type is reported; ii) \emph{error}: a tool aborted due implementation issues; for example, \verismartnq\xspace and \smartestnq\xspace only process contracts matching pre-defined templates; \verisolnq\xspace and \invconnq\xspace are not compatible with Solidity compilers $\geqslant$ 0.7.x; iii) \emph{timeout}: a tool failed to produce any results within a 6-hour time budget. Symbolic execution tools \mythrilnq\xspace, \manticorenq\xspace and \smartestnq\xspace had frequent timeouts. 

In terms of invariants generation ground truths, we evaluated them on a per invariant basis. That is, we manually inspected each generated invariant and considered it as accurate if it captured the correct properties without syntactical errors. We note that an accurate invariant can be trivial, meaning that correct invariants do not always prevent bugs.

\subsection{RQ1: {Effectiveness of Predicted Invariants for Bug Detection}} 
\label{subsec:RQ1}

\begin{table}[!t]
\caption{Reported bugs breakdown by type from 89,621 contracts. The last seven rows capture reported functional bugs. Reported bugs are not necessarily true positive or exploitable. \sys reported results include both bugs from the verifier and LLM reported bugs.}
\label{table:reported_bugs}
\rowcolors{2}{white}{gray!30}
\footnotesize
\setlength{\tabcolsep}{2.5pt}
\centering
\renewcommand{\arraystretch}{1.2}
\scalebox{0.73}{
\begin{tabular}{l|l|l|l|l|l|l|l}
\hline
Bug Type            & \sys   & \verisolnq & \smartestnq & \verismartnq & \mythrilnq & \slithernq & \manticorenq \\ \hline
RE                  & 9,011  & 1,591     & 0           & 0            & 1,311      & 2,533      & 901          \\
IF                  & 13,531 & 2,031     & 31,655       & 29,015       & 602        & 952        & 421          \\
AF                  & 11,009 & 905       & 10,921      & 12,548        & 648        & 0          & 421          \\
SC                  & 908    & 0         & 452         & 366          & 99         & 972        & 122          \\
EL                  & 611    & 0         & 0           & 0            & 82         & 1,200      & 34           \\
IG                  & 494    & 0         & 0           & 2            & 12         & 78         & 122          \\
IVO                 & 1,022  & 4,899      & 3,091       & 3,001        & 23         & 79         & 90           \\
PM                  & 2,651  & 5         & 2           & 0            & 0          & 0          & 0            \\
PE                  & 3,019  & 0         & 0           & 0            & 0          & 0          & 0            \\
BLF                 & 1,091  & 84        & 0           & 0            & 0          & 0          & 0            \\
IS                  & 977    & 33        & 5           & 5            & 107        & 0          & 0            \\
AV                  & 2,065  & 0         & 0           & 0            & 0          & 0          & 0            \\
CB                  & 3,192  & 0         & 0           & 0            & 0          & 0          & 0            \\
IDV                 & 1,924  & 0         & 0           & 0            & 0          & 0          & 0            \\ \hline
\textbf{Total Bugs} & 51,505 & 1,924     & 46,126      & 44,937       & 2,884      & 5,814      & 2,111        \\ \hline
\end{tabular}}
\end{table}
\begin{table}[!t]
\caption{Report on total count of non-compilable error and timeout results from 89,621 contracts. \sys result is from disabling verifier configuration. }
\label{table:error_timeouts}

\footnotesize
\setlength{\tabcolsep}{1.2pt}
\centering
\renewcommand{\arraystretch}{1.2}
\scalebox{0.8}{
\begin{tabular}{l|l|l|l|l|l|l|l}
\hline
        & \sys & \verisolnq & \smartestnq & \verismartnq & \mythrilnq & \slithernq & \manticorenq \\ \hline
Error   & 13 & 18,769  & 11,859      & 11,859       & 14,211     & 83,807    & 23,301      \\ \hline
Timeout & 0    & 0          & 31,636      & 32,825       & 72,526    & 0          & 64,209       \\ \hline
\end{tabular}}
\end{table}
\begin{table*}[!t]
\caption{Refined bug detection analysis on sampled 1,241 real-world functional buggy contracts with report on correct (TP), incorrect (FP), and missed (FN) bug alarms.
\color{red}\xmark\color{black}: a tool did not produce any results.}
\label{table:refined_bug_eval}
\rowcolors{3}{white}{gray!30}
\footnotesize
\setlength{\tabcolsep}{.8pt}
\centering
\renewcommand{\arraystretch}{0.9}
\scalebox{0.83}{
\begin{tabular}{l|lll|lll|lll|lll|lll|lll|lll}
\hline
Contracts                  & \multicolumn{3}{l|}{\textbf{\sys}} & \multicolumn{3}{l|}{\textbf{VeriSol}} & \multicolumn{3}{l|}{\textbf{SmarTest}} & \multicolumn{3}{l|}{\textbf{VeriSmart}} & \multicolumn{3}{l|}{\textbf{Mythril}} & \multicolumn{3}{l|}{\textbf{Slither}} &  \multicolumn{3}{l} {\textbf{Manticore}}    \\
                           & TP         & FP         & FN       & TP      & FP                & FN      & TP       & FP                 & FN     & TP       & FP                 & FN      & TP      & FP                & FN      & TP      & FP                & FN      & TP     & FP                 & FN     \\ \hline
hundredFinance             & 45         & 3          & 0        &         & \color{red}\xmark &         &          & \color{red}\xmark  &        &          & \color{red}\xmark  &         &         & \color{red}\xmark &         & 45       & 0                 & 5       &        & \color{red}\xmark  &        \\
sherlockYields             & 31         & 5          & 0        & 1       & 1                 & 3       &          & \color{red}\xmark  &        &          & \color{red}\xmark  &         & 3       & 1                 & 0       & 30       & 0                 & 4       &        & \color{red}\xmark  &        \\
dfxFinance & 72         & 6          & 0        & 1       & 0                 & 1       &          & \color{red}\xmark  &        &          & \color{red}\xmark  &         &         & \color{red}\xmark &         &         & \color{red}\xmark &         &        & \color{red}\xmark  &        \\
Bacon                      & 92         & 0          & 0        & 1       & 0                 & 1       &          & \color{red}\xmark  &        &          & \color{red}\xmark  &         &         & \color{red}\xmark &         &         & \color{red}\xmark &         &        & \color{red}\xmark  &        \\
AnySwap         & 91         & 0          & 0        & 1       & 0                 & 0       &          & \color{red}\xmark  &        &          & \color{red}\xmark  &         & 2       & 1                 & 5       & 9       & 1                 & 6       &        & \color{red}\xmark  &        \\
Dodo                     & 4          & 4          & 1        & 1       & 0                 & 0       & 5        & 0                  & 0      & 3        & 0                  & 0       & 1       & 0                 & 9       & 4       & 5                 & 3       &        & \color{red}\xmark  &        \\
Dao                        & 1          & 0          & 0        &         & \color{red}\xmark &         &          & \color{red}\xmark  &        &          & \color{red}\xmark  &         & 1       & 0                 & 2       & 2       & 2                 & 1       &        & \color{red}\xmark  &        \\
Bancor               & 24         & 1          & 0        &         & \color{red}\xmark &         &          & \color{red}\xmark  &        &          & \color{red}\xmark  &         & 1       & 0                 & 9       &         & \color{red}\xmark &         &        & \color{red}\xmark  &        \\
beanStalk                & 41         & 2          & 0        &         & \color{red}\xmark &         &          & \color{red}\xmark  &        &          & \color{red}\xmark  &         & 1       & 0                 & 10      & 4       & 2                 & 3       &        & \color{red}\xmark  &        \\
BeautyChain                        & 1          & 0          & 0        &         & \color{red}\xmark &         &          & \color{red}\xmark  &        &          & \color{red}\xmark  &         &         & \color{red}\xmark &         & 1       & 4                 & 9       &        & \color{red}\xmark  &        \\
Melo                       & 13         & 0          & 0        &         & \color{red}\xmark &         &          & \color{red}\xmark  &        &          & \color{red}\xmark  &         &         & \color{red}\xmark &         & 1       & 1                 & 10      & 4      & 1                  & 0      \\
NoodleFinance                  & 2          & 0          & 0        & 1       & 0                 & 1       & 0        & 1                  & 1      & 0        & 1                  & 1       & 3       & 1                 & 11      & 1       & 4                 & 8       & 9      & 0                  & 0      \\
BGLD                   & 2          & 0          & 0        & 1       & 0                 & 4       & 4        & 0                  & 0      & 4        & 0                  & 0       & 0       & 1                 & 23      & 1       & 4                 & 2       & 3      & 0                  & 0      \\
GYMNetwork                        & 1          & 0          & 0        & 2       & 0                 & 3       & 2        & 0                  & 0      & 2        & 0                  & 0       & 1       & 1                 & 20      &         & \color{red}\xmark &         & 2      & 2                  & 0      \\
eslasticSwap                       & 2          & 0          & 0        & 1       & 0                 & 5       & 9        & 0                  & 2      & 5        & 0                  & 2       & 4       & 0                 & 19      & 5       & 3                 & 12      & 5      & 0                  & 0      \\
EulerFinance               & 14         & 13         & 0        & 5       & 0                 & 13      & 1        & 0                  & 5      & 1        & 0                  & 5       & 5       & 0                 & 41      & 7       & 2                 & 1       & 6      & 1                  & 0      \\
Meter                      & 15         & 0          & 0        & 3       & 0                 & 4       & 9        & 0                  & 9      & 9        & 0                  & 9       &         & \color{red}\xmark &         & 1       & 5                 & 1       &        & \color{red}\xmark  &        \\
NXUSD                     & 23         & 4          & 0        & 1       & 0                 & 4       & 5        & 0                  & 1      & 5        & 0                  & 1       &         & \color{red}\xmark &         &         & \color{red}\xmark &         &        & \color{red}\xmark  &        \\
monoSwap                      & 19         & 1          & 0        & 8       & 0                 & 1       & 6        & 0                  & 4      & 6        & 0                  & 4       &         & \color{red}\xmark &         &         & \color{red}\xmark &         &        & \color{red}\xmark  &        \\
LIFI                   & 18         & 0          & 0        & 0       & \color{red}\xmark &         & 6        & 2                  & 5      & 4        & 2                  & 5       &         & \color{red}\xmark &         &         & \color{red}\xmark &         &        & \color{red}\xmark  &        \\
MuBank                        & 14         & 0          & 0        & 9       & 0                 & 0       & 0        & 1                  & 0      & 0        & 1                  & 0       &         & \color{red}\xmark &         & 9       & 5                 & 5       &        & \color{red}\xmark  &        \\
OneRing                        & 12         & 0          & 0        & 5       & 0                 & 0       & 0        & 5                  & 0      & 0        & 5                  & 0       &         & \color{red}\xmark &         & 2       & 4                 & 9       &        & \color{red}\xmark  &        \\
Paraluni                    & 6          & 0          & 0        & 6       & 0                 & 0       & 0        & 1                  & 0      & 0        & 1                  & 0       &         & \color{red}\xmark &         & 1       & 1                 & 8       &        & \color{red}\xmark  &        \\
InverseFinance             & 9          & 1          & 0        & 4       & 0                 & 2       & 0        & 1                  & 0      & 0        & 1                  & 0       &         & \color{red}\xmark &         & 1       & 2                 & 4       &        & \color{red}\xmark  &        \\
nimBus         & 3          & 0          & 0        & 0       & \color{red}\xmark &         & 0        & 1                  & 0      & 0        & 1                  & 0       &         & \color{red}\xmark &         & 1       & 1                 & 10      & 9      & 0                  & 4      \\
moneyReserve               & 4          & 2          & 0        & 3       & 1                 & 2       & 0        & 1                  & 0      & 0        & 1                  & 0       &         & \color{red}\xmark &         & 1       & 0                 & 12      & 1      & 0                  & 0      \\
pancakeswap                & 18         & 7          & 0        & 7       & 1                 & 7       & 0        & 2                  & 1      & 0        & 2                  & 1       &         & \color{red}\xmark &         &         & \color{red}\xmark &         & 3      & 1                  & 1      \\
uniswap                    & 42         & 9          & 0        & 8       & 1                 & 9       & 1        & 9                  & 7      & 1        & 8                  & 7       &         & \color{red}\xmark &         & 3       & 1                 & 3       & 2      & 1 & 1      \\
visor                      & 31         & 12         & 0        & 7       & 0                 & 9       & 9        & 6                  & 8      & 6        & 6                  & 7       &         & \color{red}\xmark &         & 2       & 8                 & 15      &        & \color{red}\xmark  &        \\
DFX                        & 2          & 9          & 0        &         & \color{red}\xmark &         & 7        & 4                  & 4      & 0        & 4                  & 4       & 5       & 2                 & 22      &         & \color{red}\xmark &         &        & \color{red}\xmark  &        \\
Harvest                       & 9          & 5          & 0        & 4       & 0                 & 8       & 0        & 10                 & 3      & 0        & 0                  & 3       & 1       & 0                 & 4       & 0       & 3                 & 1       &        & \color{red}\xmark  &        \\
moon                       & 13         & 3          & 0        &         & 0                 & 7       & 2        & 0                  & 9      & 2        & 0                  & 9       & 1       & 0                 & 19      & 1       & 2                 & 4       &        & \color{red}\xmark  &        \\
VFT                  & 2          & 2          & 0        &         & \color{red}\xmark &         & 10       & 1                  & 2      & 10       & 1                  & 2       & 1       & 0                 & 31      &         & \color{red}\xmark &         &        & \color{red}\xmark  &        \\
proxyTransfer              & 4          & 1          & 0        & 0       & 1                 & 1       & 1        & 0                  & 8      & 1        & 0                  & 8       & 1       & 0                 & 6       & 1       & 0                 & 13      &        & \color{red}\xmark  &        \\
Nomad                      & 5          & 1          & 0        & 1       & 0                 & 3       & 0        & 0                  & 1      & 0        & 0                  & 1       & 1       & 3                 & 8       & 3       & 1                 & 12      &        & \color{red}\xmark  &        \\
Fundstransfer              & 15         & 1          & 0        & 0       & 1                 & 3       & 2        & 9                  & 3      & 2        & 9                  & 3       & 1       & 1                 & 9       & 2       & 2                 & 2       & 1      & 3                  & 3      \\
walnutFinance              & 23         & 1          & 1        & 0       & 1                 & 3       & 3        & 0                  & 1      & 3        & 0                  & 1       & 0       & 0                 & 1       & 5       & 2                 & 5       & 1      & 2                  & 5      \\
Umbrella                        & 19         & 1          & 0        & 1       & 0                 & 3       & 5        & 0                  & 0      & 5        & 0                  & 0       & 1       & 1                 & 5       & 9       & 2                 & 4       & 1      & 1                  & 10      \\
Fortress Loan              & 5          & 2          & 0        & 1       & 0                 & 3       & 1        & 8                  & 0      & 1        & 8                  & 0       & 1       & 0                 & 7       & 0       & 2                 & 3       & 1      & 2                 & 3     \\
ShadowFinance                      & 4          & 0          & 0        &         & \color{red}\xmark &         & 1        & 0                  & 0      & 1        & 0                  & 0       & 1       & 0                 & 1       &         & \color{red}\xmark &         & 1      & 1                 & 3     \\
FeiProtocol                & 9          & 9          & 0        &         & \color{red}\xmark &         & 1        & 0                  & 0      & 1        & 0                  & 0       & 1       & 0                 & 5       &         & \color{red}\xmark &         & 1      & 7                  & 1      \\
Revest                     & 10         & 3          & 0        &         & \color{red}\xmark &         & 1        & 0                  & 0      & 1        & 0                  & 0       & 1       & 0                 & 9       &         & \color{red}\xmark &         & 1      & 7                  & 1      \\
Cartel                     & 3          & 0          & 0        &         & \color{red}\xmark &         & 1        & 0                  & 0      & 1        & 0                  & 0       & 1       & 0                 & 13      & 2       & 7                 & 3       & 1      & 8                  & 3      \\
Qubit                      & 11         & 2          & 0        &         & \color{red}\xmark &         & 1        & 0                  & 0      & 1        & 0                  & 0       & 1       & 0                 & 11      & 1       & 2                 & 8       & 1      & 2                  & 5      \\
ValueVaults                        & 2          & 1          & 0        &         & \color{red}\xmark &         &          & \color{red}\xmark  &        &          & \color{red}\xmark  &         &         & \color{red}\xmark &         & 1       & 1                 & 3       &        & \color{red}\xmark  &        \\
PancakeBunny               & 3          & 1          & 0        &         & \color{red}\xmark &         &          & \color{red}\xmark  &        &          & \color{red}\xmark  &         & 2       & 0                 & 5       & 1       & 0                 & 3       &        & \color{red}\xmark  &        \\
Nomad                       & 13         & 2          & 0        &         & \color{red}\xmark &         &          & \color{red}\xmark  &        &          & \color{red}\xmark  &         &         & \color{red}\xmark &         & 1       & 0                 & 5       &        & \color{red}\xmark  &        \\
SandleFinance               & 25         & 1          & 0        & 0       & 1                 & 5       & 5        & 6                  & 2      & 5        & 6                  & 2       & 0       & 4                 & 9       & 1       & 0                 & 1       &        & \color{red}\xmark  &        \\
bunnyswap                  & 16         & 1          & 0        & 2       & 0                 & 6       & 2        & 1                  & 1      & 2        & 1                  & 1       & 1       & 3                 & 10      & 1       & 0                 & 0       &        & \color{red}\xmark  &        \\
MonoX                      & 13         & 1          & 0        & 3       & 0                 & 3       & 10       & 2                  & 0      & 10       & 2                  & 0       & 6       & 2                 & 19      &         & \color{red}\xmark & 5       &        & \color{red}\xmark  &        \\
CreamFinance               & 52         & 0          & 0        & 0       & 4                 & 6       & 9        & 9                  & 1      & 9        & 9                  & 1       & 1       & 1                 & 3       &         & \color{red}\xmark & 9       &        & \color{red}\xmark  &        \\
Jay                     & 12         & 2          & 0        & 1       & 0                 & 9       & 0        & 4                  & 1      & 0        & 4                  & 1       & 9       & 2                 & 12      &         & \color{red}\xmark & 7       & 1      & 3                  & 1      \\
sushiSwap                  & 32         & 1          & 0        &         & \color{red}\xmark &         & 0        & 5                  & 3      & 0        & 5                  & 3       & 1       & 1                 & 4       & 3       & 4                 & 2       & 0      & 5                  & 3      \\
polynetwork                & 40         & 2          & 0        &         & \color{red}\xmark &         &          & \color{red}\xmark  &        &          & \color{red}\xmark  &         & 5       & 5                 & 5       & 2       & 5                 & 0       & 0      & 2                  & 4      \\
ChainSwap                  & 18         & 3          & 0        &         & \color{red}\xmark &         & 2        & 9                  & 4      & 2        & 9                  & 4       & 4       & 9                 & 9       & 3       & 6                 & 5       &        & \color{red}\xmark  &        \\
grimFinance                        & 22         & 1          & 0        &         & \color{red}\xmark &         & 9        & 1                  & 1      & 9        & 1                  & 1       & 1       & 0                 & 14      & 1       & 5                 & 2       &        & \color{red}\xmark  &        \\
Ragnarok                & 15         & 3          & 0        & 0       & 1                 & 3       & 4        & 0                  & 2      & 4        & 0                  & 2       & 1       & 1                 & 4       & 1       & 0                 & 13      &        & \color{red}\xmark  &        \\
XSurge                        & 41         & 0          & 0        & 5       & 0                 & 1       & 6        & 1                  & 1      & 6        & 1                  & 1       & 3       & 5                 & 13      & 4       & 0                 & 0       &        & \color{red}\xmark  &        \\
templeDao                  & 14         & 0          & 0        & 6       & 0                 & 0       &          & \color{red}\xmark  &        &          & \color{red}\xmark  &         & 8       & 2                 & 2       & 5       & 0                 & 13      &        & \color{red}\xmark  &        \\
RariFinance                     & 4          & 0          & 0        & 0       & 1                 & 0       &          & \color{red}\xmark  &        &          & \color{red}\xmark  &         & 7       & 2                 & 3       &         & \color{red}\xmark &         &        & \color{red}\xmark  &        \\
BabySwap                  & 5          & 0          & 0        &         & \color{red}\xmark &         &          & \color{red}\xmark  &        &          & \color{red}\xmark  &         & 2       & 1                 & 2       & 0       & 1                 & 3       &        & \color{red}\xmark  &        \\ \hline
\textbf{1241 contracts}    & 1111       & 129        & 2        & 100     & 14                & 133     & 140      & 100                & 90     & 122      & 89                 & 89      & 91      & 50                & 414     & 179     & 101               & 257     & 54     & 49                 & 48    \\
\end{tabular}}
\end{table*}

\begin{figure}[!t]

\centering
\caption{Comparison of \sys with three prompting-based tools using the 1,241 contracts in the refined experiments of Table~\ref{table:refined_bug_eval}.}
\includegraphics[width=0.9\linewidth]{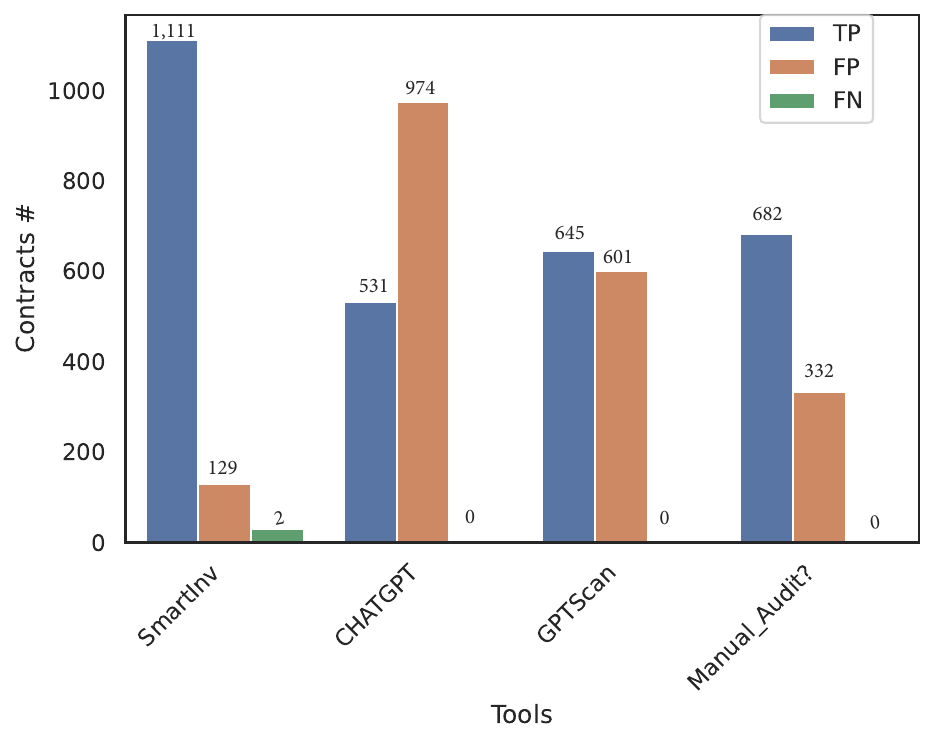}
\label{fig:promts_eval}

\end{figure}
We analyze \sys's bug-detection effectiveness by comparing it with six similar state-of-the-art tools: i) \verisolnq (as is); ii) \verismart, a CEGIS-style verifier; iii) \smartest, a language-model guided symbolic execution tool; iv) \mythril, a commercial symbolic execution tool;  v) \manticore, a commercial symbolic execution tool; vi) \slither, a static analyzer. We also compare \sys with three other prompting based approaches \cite{chatgpthowfar, gptscan, manualaudit}. Since \cite{chatgpthowfar, manualaudit} do not have a named tool, we refer to \cite{chatgpthowfar} by their model in use as \chatgpt\xspace and refer to \cite{manualaudit} by its abbreviated paper title as \ma. We refer to \cite{gptscan} by its tool \gptscan.

We conducted both large-scale and refined analysis. In our large-scale experiment, we ran each tool on the entire dataset of 89,621 contracts until a tool terminated or timed out after 6 hours.
We recorded the reported results by each bug type.
In our refined experiment, we sampled 1241 contracts from 60 hacked live dApp projects between August 1, 2022 and July 1, 2023. To provide refined false positive and false negative analysis, we confirmed a total of 1,231 bugs from audit reports for the refined analysis.

\vspace{.1cm}\noindent\textbf{Reported Bugs.} Table~\ref{table:reported_bugs} shows that \sys, \verisolnq (as is), \smartestnq, \verismartnq,  \mythrilnq, \slithernq, \manticorenq\xspace reported bugs of 57.47\%, 10.65\%, 51.47\%, 50.14\%, 3.22\%, 6.49\% on evaluated contracts. Compared to existing tools, \sys found more bugs on 5,397 contracts with major performance gains from functional bugs, identifying 14,797 more functional bugs than existing tools.

Table~\ref{table:error_timeouts} summarizes the number of errors and timeouts of each tool. The results of \smartestnq, \verismartnq, \mythrilnq, and \manticorenq\xspace demonstrate limited scalability as a major drawback of symbolic execution tools. For tools employing symbolic execution, they had timeouts on at least 35\% of evaluation contracts with a generous time budget of 6 hours. Evaluated verifiers and static analysis tool\cite{verismart, verisol, slither} have significant higher (up to 93.51\% of evaluation contracts) errors due to incompatible Solidity versions. By comparison, \sys has the lowest errors among evaluated tools, because the model is capable of reasoning contracts without requiring compiled source code.

\noindent\textbf{False Positive Analysis.} Table~\ref{table:refined_bug_eval} summarizes refined bug detection reports. \sys, \verisolnq, \smartestnq, \verismartnq, \mythrilnq, \slithernq, \manticorenq\xspace had 10.39\%, 12.28\%, 41.67\%, 42.18\%, 35.46\%, 36.07\%, 47.57\% false positive rate on our contracts. \sys was able to analyze 85\% more contracts than existing tools (up to 1,057 more contracts), because the latter had timeouts and errors.

Since existing tools relied on pattern matching for bug detection, we observed that the false positives of existing tools were largely due to matching of spurious patterns. For example, \smartestnq, \mythrilnq, \manticorenq, and \verismartnq\xspace mistakenly consider line 18 in the motivating example Listing~\ref{example3} as an integer overflow (IF) bug.

\sys's false positives resulted from bugs outside \sys's detection scope. For example, on Fei Protocol contracts, \sys's false positives were due to nine bugs arising from manipulated function selector hashing. \sys mistakenly recognized these bugs as arithmetic flaws (AF) instead of precise hashing errors. For example, the buggy pattern in Listing~\ref{false_pattern} shows a common false positive result by \sys. The first two functions \texttt{A (uint x)} and \texttt{A (bytes32 x)} have function selectors \cite{funcselector}: 0x2fbebd38 and 0xb42e8758, which should be the case for different functions. The bug lies in the third function \texttt{A (uint x,  uint y)}, which shares the same selector as \texttt{A (uint x)}. When external accounts call them, clashing selectors can cause wrongly updated accounts. \sys's false positives arose from not recognizing such clashing.
\begin{scriptsize}
\mycode
\begin{lstlisting}[float,floatplacement=H,basicstyle=\fontsize{7}{8}\selectfont, caption=\sys's false positive patterns, label={false_pattern},numbers=left, xleftmargin=2em]
function A(uint x) public returns (uint) {
        return x + 2;
    }
function A(bytes32 x) public returns (bytes32) {
        return keccak256(abi.encodePacked(x));
    }
function A(uint x, uint y) public returns (uint) {
        return x*y;
    }

\end{lstlisting}
\end{scriptsize}

\vspace{.1cm}\noindent\textbf{False Negative Analysis.} \sys reported the lowest false negative rate, with only two missed bug and false negative rate of 0.3\%. By comparison, existing tools' results had false negative rate ranging from 6.20\% to 33.36\%. Existing tools mistook the majority of contracts containing functional bugs as negative (healthy contracts). For example, none of the existing tools could catch the price manipulation (PM) bug in the Visor Finance contract in Listing~\ref{listing2}.

\vspace{.1cm}\noindent\textbf{Comparisons with Prompting Based Tools.} Since \chatgpt\xspace and \ma\xspace do not offer open source implementation, we prompted the model by strictly following the format in the papers. As for \gptscan, we used the prompting templates in its git repositories. Figure~\ref{fig:promts_eval} demonstrates \sys' effectiveness in detecting true positive bugs while minimizing false positives. \chatgpt, \gptscan, and \ma\xspace assume contracts under test are buggy, therefore they did not report negative and thus no false negative results.

Overall, \sys found 2$\times$,  1.5$\times$, and 1.5$\times$ more bugs than \chatgpt, \gptscan, \ma\xspace respectively. This gain was driven by \sys' ability to detect functional bugs, as \sys detected by up to 216 more functional bugs than the others.

Meanwhile, \sys reported the lowest false positive rate of 10.39\%, whereas \chatgpt, \gptscan, and \ma\xspace reported 7.4$\times$,  3$\times$, 2.2$\times$ more false positives respectively. This result shows the drawbacks of prompting alone: without targeted finetuning and verification, the model hallucinates more often, leading to higher false results.

\subsection{RQ2: Invariants Detection Accuracy} 
\label{subsec:RQ2}

\begin{table}[!t]
\footnotesize
\centering
\setlength{\tabcolsep}{5pt}
\renewcommand{\arraystretch}{1.1}

\caption{Mean Invariants detection Analysis. LOC (Avg.): average lines of code analyzed by a tool. \#invariants/contract: average invariants generated per contract.  \#FP/contract: average incorrect invariants generated per contract.}
\label{table:inv_table}

\begin{tabular}{llll}
\toprule
         & \sys   & \invconnq                       & \verismartnq    \\ \midrule
LOC (Avg.)                  & 1,621   & 862                            & 354          \\        
\# Invariants / Contract & 6.00    & 11.70                              & 3.00          \\ 
\#FP / Contract     & 0.32      & 5.41 
& 0.92\\ 
\bottomrule
\end{tabular}
\end{table}

\vspace{.1cm}\noindent\textbf{Invariants Inference Results.} We compared \sys's invariants detection capability with \invconnq, a Daikon-adapted smart contract invariants detector, and \verismartnq, a CEGIS-style verifier. We evaluated \invconnq\xspace and \verismartnq\xspace, because they were open-source and could be patched to generate invariants. We did not include \cider, a similar learning-based tool, because its dependencies were outdated without replacement. Table~\ref{table:inv_table} shows that \sys outperforms \invconnq\xspace and \verismartnq\xspace by 41.91\% and 64.00\% in inferring bug-critical invariants. On average, \sys analyzed 88.05\% and 357.90\% more lines of code than \invconnq\xspace and \verismartnq\xspace per contract, because existing tools had errors and timeouts on 1,102 more medium contracts and 967 more large contracts than \sys.

\vspace{.1cm}\noindent\textbf{False Positive Invariants Analysis.} Each incorrect invariant was counted as a false positive invariant. \sys's overall pattern of false positive invariants was inferring correct invariants with respect to contracts at large but at wrong program points. \invconnq's\xspace results illustrated the downside of mapping invariants detectors designed for other programming langauges (Java) to Solidity. \invconnq's\xspace false positive patterns were invariants irrelevant to smart contract programs and invariants causing compilation errors.  \verismartnq's\xspace false positive invariants pattern was that they gave false warnings to integer overflow/underflow bugs.

In terms of the number and quality of invariants inferred,  \sys inferred (up to 10$\times$) more valid invariants than the state-of-the-arts. Table~\ref{table:inv_table} shows that \sys captured fewer but less trivial invariants compared to \invconnq\xspace. For example, \invconnq\xspace inferred existing invariants in the contract source code, \eg if the source code contained \texttt{from == orig(from)}, \invconnq\xspace output \texttt{from == orig(from)} as an invariant. By comparison, \sys could reason about algebraic and loop invariants not explicitly stated in contracts. Although \verismartnq\xspace was highly effective in generating invariants against integer overflow/underflow (IF) bugs, \verismartnq\xspace could not reason about invariants from multimodal information. That is why \verismartnq\xspace produced more false positive invariants on contracts.

\subsection{RQ3: Ablation Study} 
\label{subsec:ablation}

\begin{table}[!t]
\footnotesize
\setlength{\tabcolsep}{8pt}
\centering
\renewcommand{\arraystretch}{1.1}
\caption{Finetuned candidate model evaluation (*GPT4 results were obtained from prompt engneering alone without finetuning on curated training dataset due to close 
 source).}
\label{table:model_select}
\rowcolors{2}{white}{gray!30}

\begin{tabular}{lllll}
\toprule
Model & Acc. & Prec. & Rec.  & F1 \\
 \midrule
 Alpaca \cite{Alpaca} & 0.72  & 0.72 & 0.58 & 0.65\\
 T5-Small \cite{T5} & 0.81  & 0.81 & 0.70  & 0.75\\
 GPT2 \cite{GPT2} & 0.57  & 0.57  & 0.20  & 0.30\\
GPT4* \cite{gpt4} & 0.44  & 0.43  & 0.32  & 0.19\\
OPT-350M \cite{OPT} & 0.53 & 0.53 & 0.25  & 0.34\\
\midrule
LLaMA-7B \cite{LLaMA} & \textbf{0.89} & \textbf{0.89} & \textbf{0.83}  & \textbf{0.82}\\
\bottomrule
\end{tabular}
\end{table}

We considered six foundation models as our baselines: OPT-350M \cite{OPT}, Google's T5-Small~\cite{T5}, OpenAI's GPT2~\cite{GPT2} and GPT4~\cite{gpt4}, Stanford's Alpaca~\cite{Alpaca}, and Meta's LLaMA-7B~\cite{LLaMA}. Notably, GPT4 was not available for finetuning on customized datasets (as of July 31, 2023), so we supplied test contract source code and applied ToT prompting to GPT4 without finetuning. We finetuned the remaining five candidate models with multimodal information (without architectural optimization). 

We quantified how much our key optimization strategies improved the end results. 
To evaluate each of \sys's strategies, we removed natural language modality in the dataset by deleting implementation-related comments and renaming function and variable names without giving away domain-specific information, \eg we renamed ``votingToken" variable to ``var." We removed ToT by using a one-shot general prompt as ``What are the vulnerabilities and invariants in the contract?'' As for labeled features, we removed all labeled features in the training dataset. For optimization, we compared \sys' performance on unmodified LLaMA-7B against the architecturally optimized model.

\begin{table}[!t]
\footnotesize
\setlength{\tabcolsep}{6pt}
\centering
\renewcommand{\arraystretch}{1.1}
\caption{Ablation study. Natural Language: natural language modality. Labelled Features: labeled features in training data. ToT: tier of thought prompting. Optimization: model architectural optimization. Full \sys: no strategy removed and \sys is at the default setting of source code and natural language modalities. Full \sys with Tx. Hist.: ``Tx. Hist.'' refers to deployed contracts' transaction history. \sys with Tx. Hist. is finetuned on source code, natural language, transaction history modalities. }
\label{table:strategy_eval}
\rowcolors{2}{white}{gray!30}
\begin{tabular}{lllll}
\toprule
Remove & Acc. & Prec. & Rec.  & F1 \\
\midrule
All & 0.12  & 0.15  & 0.10 &0.14 \\
Natural Language & 0.62 & 0.60 & 0.30 & 0.45\\
Labeled Features & 0.59 & 0.61 & 0.60 & 0.60\\
ToT & 0.24  & 0.18  & 0.20 &0.16\\
Optimization & 0.89 & 0.88 & \textbf{0.85} & 0.82\\
\midrule
Full \sys & \textbf{0.89}  &  \textbf{0.89}  & 0.83 & \textbf{0.82}\\
Full \sys with Tx. Hist. & 0.89  &  \textbf{0.92} & \textbf{0.84} & \textbf{0.85}\\
\bottomrule
\end{tabular}
\end{table}

\vspace{.1cm}\noindent\textbf{Model Selection Results.} Table~\ref{table:model_select} demonstrates that finetuned LLaMA outperformed the second best candidate T5-Small by 8\% in accuracy and precision, 13\% in recall, and 7\% in F1. 
This observation implies scaling law~\cite{wei2022emergent} also applies to smart contract invariant inference. 

\vspace{.1cm}\noindent\textbf{Effect of Natural Language Modality.} Table~\ref{table:strategy_eval} shows that once we removed natural language modality, the accuracy and F1 of \sys dropped by 27\% and 37\% respectively. The loss in F1 was related to recall. \sys's recall dropped from 83\% to 30\% once we removed natural language. We also examined the effect of natural language modality on a per-contract basis. We observed that both \sys with natural language information and without it detected implementation bugs equally well. However, \sys with natural language modality detected far ($40\times$) more functional bugs than single-modal \sys, with major performance gain from natural language modality. 

\vspace{.1cm}\noindent\textbf{Effect of labeled Features.} Without labeled features, \sys's accuracy, precision, recall and F1 dropped by 30\%, 28\%, 23\%, and 22\% respectively. After removing labeled features, \sys lost the ability to precisely locate potentially vulnerable program points and thus predicted trivial invariants at incorrect prorgram points. 

\vspace{.1cm}\noindent\textbf{Effect of ToT.} Table~\ref{table:strategy_eval} demonstrates that ToT had a significant impact: \sys's accuracy, precision, recall, and F1 dropped by 65\%, 71\%, 63\%, and 66\% respectively when ToT was removed. Without ToT, \sys repeated the invariants in the test contracts without generating bug-preventive invariants. This result shows that ToT was crucial to \sys's bug detection performance.

\vspace{.1cm}\noindent\textbf{Effect of Quantization and PEFT.} Table~\ref{table:strategy_eval} shows architecturally optimized \sys achieved comparable results as \sys without architectural optimization. 
\sys with optimization and that without had the same accuracy of 89\%. 
While \sys with optimization outperformed that without by 1\% in precision, the latter outperformed the former by 2\% in recall. 
Both had the same F1 score of 94\%, indicating that Quantization and PEFT did not drastically increase false positives or false negatives. 
This result shows that architecturally optimized \sys decreased the cost of finetuing computation and memory by 75\% without incurring significant accuracy loss. 

\vspace{.1cm}{\noindent\textbf{Transaction History as Additional Modality.} We also conducted additional experiments on using transaction history as an additional modality for deployed contracts. We incorporated transaction history that covered a wide range of bugs as shown in Table~\ref{bugtab} during finetuning and inference. Specifically, for finetuning, we crawled the transaction history of 329 deployed contracts from Etherscan and added them in the corresponding contracts in our training data. The included transaction history information were users' addresses, call functions, and call data. For inference, we tested \sys with transaction history on previously unseen deployed contracts in our evaluation dataset.

We modified prompt B in Tier 1 as ``Given [transaction history] and transactional context, what are the critical program points?'' With transaction history, \sys's precision and F1 score in Table~\ref{table:strategy_eval} improves by 2.9 \% and 3\% respectively. This improvement applies universally to a broad range of functional bugs. Transaction history can be helpful for \sys to focus on critical and bug-prone functions. Given that \sys's key strength is to detect bugs in smart contracts' source code pre-deployment, \sys does not assume available transaction history modality by default. 
However, we expect that reasoning about transaction history can further improve \sys's bug detection performance on deployed contracts.

\subsection{RQ4: Runtime Performance}
\label{subec:runtime}
\begin{table}[!t]
\footnotesize
\centering
\setlength{\tabcolsep}{5pt}
\renewcommand{\arraystretch}{1.1}
\rowcolors{3}{white}{gray!30}
\caption{Mean Runtime Analysis (in Seconds) of Each Tool on Evaluation Dataset.}
\label{table:runtime_table}

\begin{tabular}{llllr}
\hline
                   & Small  & Medium & Large   & \multicolumn{1}{l}{Full} \\
\#Contracts        & 65,739 & 12,011 & 11,871  & 89,621                    \\ \hline
\sys       & 15.02  & 32.98  & 37.77   & 28.59                     \\
VeriSol  & 232.51  & 1612.32   & 3933.21    & 2994.01                      \\
SmarTest  & 175.01  & 297.02  & 3908.32  & 2793.45                     \\
VeriSmart & 27.21  & 33.76  & 4145.22   & 3105.40                     \\
Mythril   & 404.98 & 305.22 & 5031.33 & 3580.51                    \\
Slither   & 22.35  & 155.41  & 9080.62   & 3451.13                     \\
Manticore & 301.33 & 562.21 & 7281.94 & 4715.16                    \\ \hline
\end{tabular}
\end{table}

 we ran each tool on the entire evaluation dataset and recorded the runtime accordingly. Before presenting the results, we note that training time is amortized into inference runtime. Since \verisolnq\xspace requires manual specification, we thus approximated the manual effort of \verisolnq\xspace by consulting multiple smart contract audit companies \cite{chainlink, cypher}. We learned that each real-world contract takes an experienced auditor from 1 hour to 3 hours. We averaged 90 minutes per contract as manual specification estimate for \verisolnq.

Table~\ref{table:runtime_table} shows the average runtime on a per contract basis. \sys outperforms static tools (\verismartnq, \verisolnq, \slithernq) by up to 125$\times$ and outperforms dynamic tools (\smartestnq, \manticorenq, \mythrilnq) by up to 163$\times$. Notably, \sys's runtime overhead does not increase by more than 17 seconds on average when the size of contracts increases by 500 lines of code. As an average speedup of \sys on the entire evaluated dataset, \sys reduces overhead by up to 150$\times$.

\section{Limitations}
\label{sec:discussions}

\vspace{.1cm}\noindent\textbf{Token Length.} Foundation models have limited token length available for finetuning. LLaMA models are limited to 4096 tokens (approximately 3000 words). Therefore, large contracts with more than 2000 lines are often cut short in the reasoning process. We made initial steps towards remedying token length limitations by prompting the model to summarize imported modules in a contract under test. For future work, we plan to incorporate other promising strategies such as retrieval augmented prompting \cite{retrievalaugprompting}.

\vspace{.1cm}\noindent\textbf{Verifier Compatibility with Solidity Compilers.} We design our verifier based on \verisolsnq\xspace mappings between solidity and Boogie. \verisolnq\xspace is limited to solidity compiler between 0.4.0 and 0.7.0. Our verifier also inherits the limitation of \verisolsnq\xspace. We acknowledge that a large number of contracts do not have compatible compiler versions. In that case, we manually reviewed \sys inferred invariants and vulnerabilities. As future work, we plan to expand our verifier across newer solidity compiler versions.

\vspace{.1cm}\noindent\textbf{Exploitabiltiy of Zero-Day Bugs.} We acknowledge that not all detected zero-day bugs are exploitable. For instance, integer overflow/underflow (IF) bugs only exist in contracts built on older solidity compilers ($\leqslant$ 0.6.0). New solidity compilers automatically checks for over/underflow and thus preempt such exploitability. Newly upgraded proxy contracts also prevent exploitable zero day bugs.

\vspace{.1cm}\noindent\textbf{Threats to Validity.} Our results were obtained on our evaluation dataset, which might not be representative of newer contracts. Secondly, we did not report results based on bugs' exploitability and did not compare \sys with other tools in that regard. Evaluation based on exploitable bugs may be different. Thirdly, despite of our best effort to be precise and accurate, manual inspection on \sys's inferred vulnerabilities and manual classification of reported results into true and false positives are inherently challenging and can be subjective in some cases. 

\vspace{.1cm}\noindent\textbf{Ethical Disclosure.} 

\section{Case Study: Zero-Day Bugs}
\label{subsec:zeroday}
To illustrate \sys -detected 119 zero-day bugs, this section provides two developer-confirmed examples. We reported the zero-day bugs and were rewarded bounty of \$17,600. We anonymized the contract in \S\ref{subsec:cross-bridge} at request.

\subsection{Cross Bridge}
\label{subsec:cross-bridge}
Listing~\ref{0day1} provides a buggy code snippet of a cross bridge contract. \sys discovers the zero-day vulnerability related to the assumed unique \texttt{\_msgHash} at line 20. Since the default value of unknown \texttt{\_msgHash} is \texttt{0x00} in Solidity, this bug can potentially greenlight a malicious actor's \texttt{\_msgHash} value that defaults to \texttt{0x00} during cross bridge communication. As a result, the malicious actor can bypass the assertion check at line 20, where \texttt{messages[0x00]} is an acceptable root. \sys detects this vulnerability with invariant \texttt{(assert(\_msgHash != 0)} to prevent incorrect default values.

\mycode
\begin{lstlisting}[float,floatplacement=H,basicstyle=\fontsize{7}{8}\selectfont, caption={anonymized contract code snippet}, label={0day1}, numbers=left, xleftmargin=2em]
contract Bridge {
  function init(
    uint32 _callSite,
    address _sender,
    bytes32 _merkleRoot
     ) public {
        base_initialize(_sender);
        callSite = _callSite;
        committedRoot = _merkleRoot;
        //invariant #1: assert(_merkleRoot != 0);
        confirmAt[_merkleRoot] = 1;
     }
...

 function process(bytes memory _message) 
    public returns (bool _success) {
    ...
    //zero day vulnerability
    //invariant #2: assert(_msgHash != 0);
    assert(accept(messages[_msgHash]));
    }
    
  function accept(bytes32 _root) 
    public view returns (bool) {
    //invariant #3: assert(_root != 0);
    uint256 _time = confirmAt[_root];
     }

\end{lstlisting}

\subsection{Inefficient Gas} 

Listing \ref{0day2} contains a gas inefficient \texttt{remove()} function that can lock users' funds. A user can join the vault by first joining \texttt{DepositQueue}. While on the queue, users can choose to refund their deposit or to process it at end of a transaction round. 
However, line 7 uses a gas-expensive for-loop implementation to remove each deposit on the queue. 
A long queue can easily exceed the 30 million per block gas limit even with a single deposit operation \cite{vault}. 
An attacker can send funds from different accounts to occupy the queue. 
As a result, the contract will lose the ability to refund or process users' deposits, because all deposits are locked in \texttt{DepositQueue} due to insufficient gas.

\mycode
\begin{lstlisting}[float,floatplacement=H,basicstyle=\fontsize{7}{8}\selectfont, caption={code snippets of baseVault.sol}, label={0day2}, numbers=left, xleftmargin=2em]
abstract contract BaseVault {

    DepositQueueLib DepositQueue;

    function processQueuedDeposits(uint256 startIndex, uint256 endIndex) external {
        uint256 _totalAssets = totalAssets();
        for (uint256 i = startIndex; i < endIndex; i++){
            uint256 currentAssets = _totalAssets + processedDeposits;
            depositEntry = depositQueue.get(i);
            processedDeposits += depositEntry.amount;
        }
        //invariant #1: require(depositQueue.size()==1, "Cannot process multiple deposits");
        depositQueue.remove(startIndex, endIndex);
    }
   
\end{lstlisting}

\sys discovers this bug by generating an invariant after line 12 to check the size of \texttt{DepositQueue}. The inferred invariant \texttt{require(depositQueue.size() == 1,  "Cannot process multiple deposits")} ensures that there is only one deposit on the queue each time. This specified property is important for safeguarding the \texttt{remove()} function from running out of gas.

\section{Related Work}
\vspace{.1cm}\noindent\textbf{Smart Contract Static and Dynamic Analysis.} \smartest, \mythril, \manticore, \maian, \teether, and \ethbmc\xspace are symbolic execution tools that generate vulnerable transaction sequences. \smartestnq\xspace utilizes language models to supplement symbolic execution on smart contracts. \maiannq\xspace and \teethernq\xspace focus on high-level bugs such as Ether-leaking and suicidal vulnerabilities. \ethbmcnq\xspace focuses on memory modeling and cryptographic hashing. They largely rely on pattern specific heuristics to detect certain classes of implementation bugs.

\slither, \oyente, \osiris, and \honeybadger\xspace are static analysis tools that utilize data flow analysis. They analyze the source code of a smart contract to identify potential security vulnerabilities. Static analysis cannot reason about mutlimodal input. As a result, they are limited in their abilities to detect functional bugs.

Fuzzing is also common dynamic analysis used by smart contract security researchers. \confuzzius, \echidna, \fluffy\xspace and \sfuzz\xspace are recently developed smart contract fuzzers. They send random inputs to a smart contract and try to trigger unexpected behavior and identify potential security vulnerabilities. Fuzzing-based tools tend to be slow, because they need to explore many possible transactional states of a contract.

\vspace{.1cm}\noindent\textbf{Invariants Detectors and Verifiers.} One popular program analysis approach is invariants detection. \daikon\xspace and \invcon\xspace are both invariants detection tools. Daikon does not apply to Solidity and \invconnq\xspace is a \daikonnq-adapted tool that maps Solidity to Java. \smtchecker, \solc, \verisol, and \zeus\xspace are verification frameworks that require manually specifying invariants first and then automatically infer transaction invariants during verification. Manual specification is often error-prone.

\vspace{.1cm}\noindent\textbf{ML-Based Tools.} ML-based tools include \svchecker\xspace and \contractgraph. They use deep neural networks to discover limited sets of implementation bugs such as reentrancy based on general patterns. \escort\xspace is based on byte-code level transfer learning and \lastsys\xspace uses language models to detect code clones. Existing prompting-based tools \cite {gptscan, chatgpthowfar, manualaudit} do not check for model hallucinated results, and therefore produce many false positives. Prior ML-based tools cannot reliably detect critical functional bugs, such as flashloan-based price manipulation and privilege escalation, because they rely bug-specific graph search heuristics.

\section{Conclusion}

This paper introduced \sys, an automated framework for detecting both implementation and functional bugs in smart contracts. 
\sys can reason across multiple modalities of smart contracts, including source code and natural language, and reason over them based on a new prompting strategy called Tier of Thoughts (TOT) to generate bug-preventive invariants. Evaluation of \sys on real-world contracts revealed \sys performed well on both invariants generation and bug detection tasks.

\section*{Acknowledgement}
We would like to express our sincere appreciation to Jianan Yao for his foundational guidance on this project and invaluable advice, as well as to Andreas Kellas, Chengzhi Mao, and Zhuo Zhang for their extensive edits and feedback. 
We also extend our gratitude to the anonymous
reviewers and our Shepherd for their constructive comments, which significantly improve this paper.
This work was supported in part by Columbia Center for Digital Finance and Technologies and gifts from Google, Accenture, and DiDi.

\bibliographystyle{plain}
\bibliography{paper}

\newpage 

\appendices 

\section{Meta-Review}

The following meta-review was prepared by the program committee for the 2024
IEEE Symposium on Security and Privacy (S\&P) as part of the review process as
detailed in the call for papers.

\subsection{Summary}
The paper presents SMARTINV, a smart contract invariant inference framework to automate the detection. The key insight of the paper is that the expected behavior of smart contracts, as specified by invariants, relies on understanding multimodal information, such as source code and natural language. Thus, SMARTINV combines the analysis of source code and natural language document to detect bugs in smart contracts. It uses a tiered prompting strategy to identify invariants by applying machine learning on source code and relevant comments and documents. These invariants are then used to detect bugs, particularly functional bugs. SMARTINV is evaluated relatively thoroughly on bug detection, invariants generation, and performance.

\subsection{Scientific Contributions}
\begin{itemize}
\item Provides a Valuable Step Forward in an Established Field.
\end{itemize}

\subsection{Reasons for Acceptance}
\begin{itemize}
\item The paper develops Tier of Thought (ToT), a general prompting strategy, that can be used to fine tune and elicit explicit reasoning of foundation models on the program structures of smart contracts.

\item SMARTINV extracts invariants for expected behaviors of smart contracts by leveraging foundation models to reason about multimodal information including source code and natural language documents.

\item The comprehensive experiments demonstrate SMARTINV's superiority over the state-of-the-art approaches in invariance inference.
\end{itemize}

\end{document}